\documentclass[]{pasj01}
\usepackage[switch,mathlines]{lineno}

\Received{}
\Accepted{}
 
\begin{document} 

\title{ 
Dynamical properties of mildly relativistic ejecta produced by the mass-loading of gamma-ray burst jets in dense ambient media}


\author{Akihiro \textsc{Suzuki}\altaffilmark{1}}%
\altaffiltext{1}{Research Center for the Early Universe, The University of Tokyo, 7-3-1 Hongo, Bunkyo-ku, Tokyo 181-8588, Japan}
\email{akihiro.suzuki@resceu.s.u-tokyo.ac.jp}

\author{Christopher \textsc{M. Irwin},\altaffilmark{1}}

\author{Keiichi \textsc{Maeda}\altaffilmark{2}}
\altaffiltext{2}{Department of Astronomy, Kyoto University, Kitashirakawa-Oiwake-cho, Sakyo-ku, Kyoto, 606-8502, Japan}


\maketitle

\begin{abstract}

We present the results of a series of 3D special relativistic hydrodynamic simulations of a gamma-ray burst (GRB) jet in a massive circumstellar medium (CSM) surrounding the progenitor star. 
Our simulations reproduce the jet morphology transitioning from a well-collimated state to a thermal pressure-driven state for a range of CSM masses and outer radii. 
The jet-CSM interaction redistributes the jet energy to materials expanding into a wide solid angle and results in a quasi-spherical ejecta with 4-velocities from $\Gamma\beta\simeq 0.1$ to $\simeq 10$. 
The mass and kinetic energy of the ejecta with velocities faster than $0.1c$ are typically of the order of $0.1\,M_\odot$ and $10^{51}\,\mathrm{erg}$ with only a weak dependence on the CSM mass and radius for the explored CSM parameter ranges. 
We find that the numerically obtained density structure of the mildly relativistic ejecta is remarkably universal. 
The radial density profile is well approximated as a power-law function of the radial velocity with an index of $-5$, $\rho\propto v^{-5}$, in agreement with our previous simulations and other studies, as well as those suggested from recent studies on early-phase spectra of supernovae associated with GRBs. 
Such fast ejecta rapidly becomes transparent following its expansion. 
Gradually releasing the trapped thermal photons, the ejecta gives rise to bright UV--optical emission within $\sim 1$ day. 
We discuss the potential link of the relativistic ejecta resulting from jet-CSM interaction to GRB-associated supernovae as well as fast and blue optical transients. 
\end{abstract}

\section{Introduction}
\label{sec:introduction} 
Modern multi-wavelength astronomical transient surveys have unveiled a universe filled by various kinds of transient phenomena with the brightness and the evolutionary time-scales varying by many orders of magnitude.  
Among them, gamma-ray bursts (GRBs) are still mysterious short-time-scale phenomena, in which a tremendous amount of radiation is released from a relativistic jet/outflow within only $0.1$ to thousands of seconds \citep{1999PhR...314..575P,2004RvMP...76.1143P,2006RPPh...69.2259M,2015PhR...561....1K}. 
Long-duration GRBs are widely believed to arise in the gravitational collapse of massive stars due to their association with core-collapse supernovae (CCSNe) \citep{2006ARA&A..44..507W,2013RSPTA.37120275H,2017AdAst2017E...5C}. 
Their afterglows are powered by the external shock driven by the jet in the ambient medium and outshine radiation across the electromagnetic spectrum. 
The prompt and afterglow emission properties of long GRBs vary widely  from nearby underluminous to cosmological luminous events, suggesting a diversity of jet conditions and environments (e.g., \cite{2012ApJ...750...68L}). 

The circum-burst environments of GRB progenitors are still unclear. 
Normal GRBs at cosmological distances are thought to happen in a dilute ambient medium, which allows the jet to propagate almost freely at the beginning and then dissipate its energy over a long period, leading to afterglow emission.  
On the other hand, a dense ambient medium can hinder or even suppress the jet propagation. 
For over a decade, particular attention has been paid to GRB jets in  massive circum-stellar media (CSM) as a way of explaining an under-luminous class of GRBs (i.e., low-luminosity GRBs; \cite{2006Natur.442.1008C,2006Natur.442.1011P,2006Natur.442.1014S,2006Natur.442.1018M}). 
The presence of a dense CSM around the progenitor of low-luminosity GRBs is indeed inferred from the observed early optical emission (e.g., \cite{2015ApJ...807..172N,2016MNRAS.460.1680I}). 
GRBs with unusually long duration (ultra-long GRBs; e.g., \cite{2014ApJ...781...13L}) are suggested to arise from massive stars with their envelope attached, such as blue-supergiants (BSGs) (e.g., \cite{2013ApJ...766...30G,2013ApJ...778...67N}), in which the jet must drill through the massive envelope. 
A GRB jet in a sufficiently massive CSM/extended envelope would lead to a so-called failed or choked jet due to the high ram pressure of the ambient gas. 
GRB jets suffering from such heavy mass-loading have long been hypothesized as ``dirty'' fireballs (\cite{1998ApJ...494L..45P,1999ApJ...513..656D,2002MNRAS.332..735H}) as opposed to 
``clean'' fireballs accelerating to ultra-relativistic velocities. 
It is still unclear whether some GRB progenitors have a massive CSM or an extended envelope at the onset of their gravitational collapses. 
However, recent observations of CCSNe especially in their infant phase have revealed the presence of a confined CSM around massive stars in their pre-supernova stage not only for hydrogen-rich (e.g., \cite{2017NatPh..13..510Y,2018NatAs...2..808F}) but also for hydrogen-poor/deficit progenitors (\cite{2021ApJ...918...34M,2021arXiv210807278F,2022Natur.601..201G,2022ApJ...927..180P}). 
The additional energy dissipation caused by the CSM interaction produces an early optical peak separated from the main peak powered by radioactive nickel (double-peaked SNe; 
 e.g., \cite{2013ApJ...769...67P,2014ApJ...788..193N}). 

Besides, recent optical transient surveys have started detecting GRB afterglow-like optical transients without any gamma-ray trigger \citep{2015ApJ...803L..24C,2022ApJ...938...85H}. 
About a half of them lack gamma-ray counterparts and therefore could be truly GRB-less events, i.e., either off-axis (orphan) afterglows or afterglows from dirty fireballs. 
The current observations have already put an upper limit on the rate of dirty fireballs with a kinetic energy similar to on-axis GRBs. 
Observational studies of orphan GRB afterglows in the near future would give us important constraints on the GRB jet structure and mass-loading \citep{2024arXiv240116470P}. 

Another class of optical transients possibly powered by relativistic outflows has been identified in the last decade; the fast and blue optical transients or FBOTs for short (\cite{2010ApJ...724.1396O,2014ApJ...794...23D,2016ApJ...819...35A,2016ApJ...819....5T,2018MNRAS.481..894P,2019ApJ...885...13T,2020ApJ...894...27T,2022ApJ...933L..36J}). 
Despite dedicated multi-wavelength follow-up observations of the best studied FBOT, AT 2018cow \citep{2018ApJ...865L...3P,2019MNRAS.487.2505K,2019MNRAS.484.1031P,2019ApJ...872...18M,2019ApJ...878L..25H,2020MNRAS.491.4735B,2022NatAs...6..249P}, its origin and power source(s) are still extensively debated. 
The short rise and decay time-scales of their optical emission imply a smaller ejecta mass $\sim 0.1M_\odot$ than normal CCSN explosions ejecting $>1M_\odot$ of materials, thereby requiring a mechanism to deposit a large amount of energy into a particularly small mass. 
A variety of progenitor scenarios have been proposed in the literature \citep{2019MNRAS.488.3772F,2019MNRAS.485L..83Q,2019MNRAS.487.5618L,2019ApJ...877L..21Y,2020ApJ...903...66L,2020ApJ...894....2P,2020ApJ...897..156U,2021ApJ...910...42X,2022ApJ...932...84M}.

FBOTs with especially bright UV emission are also identified as bright radio transients with their radio luminosities comparable to radio-bright SNe and GRB afterglows at late epochs  \citep{2019ApJ...871...73H,2019ApJ...872...18M,2020ApJ...895L..23C,2020ApJ...895...49H,2021ApJ...912L...9N,2022ApJ...926..112B,2022ApJ...932..116H,2022ApJ...934..104Y}. 
The analyses of radio-bright FBOTs suggest a blast-wave synchrotron origin with a required shock velocity of $\sim 0.1c$ or faster \citep{2020ApJ...895L..23C,2022ApJ...926..112B,2022ApJ...932..116H}. 
This finding led some authors to put forward scenarios with relativistic jets/outflows from stellar collapses (e.g., \cite{2021MNRAS.508.5138P,2022MNRAS.513.3810G}). 

All these lines of evidence suggest essential roles played by relativistic jets and their interaction with ambient gas in various transients. 
Even for a successful jet from a compact progenitor, a quasi-spherical or bipolar 
ejecta component, the so-called cocoon, is a natural by-product of jet penetration through the star. 
A massive CSM around a GRB progenitor potentially has an even more serious impact on the cocoon expansion \citep{2013ApJ...764L..12S} or chokes the jet \citep{2015ApJ...807..172N}. 
In general, jet injection into a massive star or dense ambient gas inevitably leads to (partial) dissipation of the jet energy, which is redistributed among the materials ejected into a wide solid angle and gives rise to a variety of electromagnetic signals (e.g., \cite{2002MNRAS.337.1349R,2017ApJ...834...28N,2018MNRAS.478.4553D,2018ApJ...863...32D,2022MNRAS.512.3627D}). 
Predicting emission powered by a GRB jet interacting with a massive CSM requires understanding of the physical properties of the ejecta resulting from jet-CSM interaction, such as the density structure and the amount of the kinetic and internal energies loaded on the ejecta. 
There are several analytical and numerical considerations of GRB jet propagation and mass-loading in a dense environment (e.g., \cite{2013ApJ...764L..12S,2015MNRAS.446.1716C,2015MNRAS.446.1737C,2015ApJ...807..172N,2019MNRAS.489.2844I,2020ApJ...900..193D,2022MNRAS.517..582E,2023MNRAS.519.1941P}). 
However, the properties of the ejecta well after the jet termination are only poorly investigated so far. 
\citet{2022ApJ...925..148S} have conducted a long-term 3D simulation of a GRB jet propagating in a $0.1M_\odot$ CSM and revealed the formation of a mildly relativistic ejecta with a characteristic radial density structure. 
However, the immediate circumstellar environment of GRB progenitors is still poorly known and a wide parameter exploration is required. 
In this study, we extend GRB jet simulations by \citet{2022ApJ...925..148S} to include new models with wider parameter sets for properties of CSM surrounding the progenitor.

This paper is organized as follows. 
In Section \ref{sec:numerical_simulations}, we describe our numerical simulation setups and present the results. In Section \ref{sec:radiation}, we summarize the ejecta properties and consider the expected emission from the ejecta. 
In Section \ref{sec:discussion}, we discuss astronomical transients possibly powered by jet-CSM interaction. 
We finally summarize our findings in Section \ref{sec:summary}. 
We provide detailed descriptions of our light curve model in Appendix. 

\section{Numerical Simulations}\label{sec:numerical_simulations}
\subsection{Setups}
We perform hydrodynamic simulations based on our previous work \citep{2022ApJ...925..148S}, in which we computed GRB jet propagation in and outside of a pre-supernova $14\,M_\odot$ CO star with the radius of $0.58\,R_\odot$.  
The progenitor model is adopted from \citet{2006ApJ...637..914W} (16TI model). 
We use a 3D special-relativistic hydrodynamics simulation code equipped with an adaptive mesh refinement (AMR) technique \citep{2017MNRAS.466.2633S,2019ApJ...880..150S}, which solves the time evolution of the density $\rho$, the three components of the velocity $(\beta_x,\beta_y,\beta_z)$ (normalized by the speed of light $c$), and the pressure $p$ in a computational domain with Cartesian coordinates $(x,y,z)$. 
The equation of state for ideal gas with the adiabatic index $\gamma=4/3$ is assumed. 
\subsubsection{Jet and energy injection}
We employ the same jet injection conditions as the standard jet model in \citet{2022ApJ...925..148S}: a (one-sided) jet energy injection rate of $L_\mathrm{jet}=2.5\times10^{50}\,\mathrm{erg}\,\mathrm{s}^{-1}$, a jet half-opening angle of $\theta_\mathrm{jet}=10^\circ$, an initial Lorentz factor of $\Gamma_\mathrm{0}=5$, and a specific energy of $\epsilon_\mathrm{0}/c^2=20$. 
\citet{2022ApJ...925..148S} have demonstrated that a jet injected in this way penetrates the star $t_\mathrm{br}\simeq 5$ s after the injection and accelerates up to a terminal Lorentz factor of $\sim100$ in the absence of a massive CSM. 
We stop the jet injection at $t=20\ $s, ending up with a total injected jet energy of $5\times10^{51}\,$erg. 
In addition to the jet injection, we also assume an instantaneous and spherical energy deposition with $5\times 10^{51}\,$erg, corresponding to the associated SN explosion.

\subsubsection{Ambient medium}

In this work, we extend the jet model of \citet{2022ApJ...925..148S} who considered a $0.1\,M_\odot$ CSM around the progenitor, to cover a wider parameter space for the CSM properties. 
We assume a wind-like CSM with spherical symmetry. 
The radial density profile is given by
\begin{equation}
    \rho_\mathrm{csm}(r)=\frac{M_\mathrm{csm}}{4\pi r^2R_\mathrm{csm}p\Gamma(1/p)}\exp\left[-\left(\frac{r}{R_\mathrm{csm}}\right)^p\right],
\end{equation}
where $\Gamma(x)$ is a gamma function and the index $p$ is set to $p=10$. 
The inner part of the CSM density can be approximated by a simple inverse square law, $\rho_\mathrm{csm}\simeq A_\mathrm{csm}r^{-2}$ with 
\begin{equation}
    A_\mathrm{csm}=\frac{M_\mathrm{csm}}{4\pi R_\mathrm{csm}p\Gamma(1/p)}=
    5.9\times10^{18}
    \left(\frac{M_\mathrm{csm}}{M_\odot}\right)
    \left(\frac{R_\mathrm{csm}}{40\,R_\odot}\right)^{-1}
    \,\mathrm{g}\,\mathrm{cm}^{-1}. 
\end{equation}
This density profile has a couple of characteristic quantities, the outer radius $R_\mathrm{csm}$ and the CSM mass $M_\mathrm{csm}$. 
The CSM density profile has a sharp cut-off around $r=R_\mathrm{csm}$. 
The density of the CSM is therefore characterized by these two parameters. 
Beyond the CSM outer radius, we put another wind-like medium;
\begin{equation}
    \rho_\mathrm{w}(r)=A_\mathrm{w}r^{-2},
\end{equation}
with a much smaller normalization constant, $A_\mathrm{w}=5.0\times10^{11}\,\mathrm{g}\,\mathrm{cm}^{-1}$, which does not affect the jet propagation.

\begin{table}
  \tbl{Model description}{%
\begin{tabular}{lrrr}
\hline\hline
Model&$M_\mathrm{csm}[M_\odot]$&$R_\mathrm{csm}[\mathrm{cm}]$&$A_\mathrm{csm}[\mathrm{g}\,\mathrm{cm}^{-1}]$
\\
\hline
M01R40&$0.1$&$2.8\times 10^{12}$&$5.9\times10^{17}$\\
M03R40&$0.3$&$2.8\times 10^{12}$&$1.8\times10^{18}$\\
M1R40&$1.0$&$2.8\times 10^{12}$&$5.9\times10^{18}$\\
M3R40&$3.0$&$2.8\times 10^{12}$&$1.8\times10^{19}$\\
M10R40&$10.0$&$2.8\times 10^{12}$&$5.9\times10^{19}$\\
M01R400&$0.1$&$2.8\times 10^{13}$&$5.9\times10^{16}$\\
M03R400&$0.3$&$2.8\times 10^{13}$&$1.8\times10^{17}$\\
M1R400&$1.0$&$2.8\times 10^{13}$&$5.9\times10^{17}$\\
M3R400&$3.0$&$2.8\times 10^{13}$&$1.8\times10^{18}$\\
M10R400&$10.0$&$2.8\times 10^{13}$&$5.9\times10^{18}$\\
\hline\hline
\end{tabular}}
\label{table:model_description}
\end{table}

We present 10 simulations with different sets of $M_\mathrm{csm}$ and $R_\mathrm{csm}$. 
Table \ref{table:model_description} summarizes the models and their parameters. 
We assume a couple of CSM radii, $2.8\times 10^{12}$ cm ($\simeq 40\,R_\odot$) and $2.8\times10^{13}$ cm ($\simeq 400\,R_\odot$). 
For each CSM radius, we conduct 5 simulations with different CSM masses from $0.1\,M_\odot$ to $10\,M_\odot$. 
We follow the evolution of the system up to $t=10^4\,$s.  

\subsection{Results}\label{sec:results}
\subsubsection{Jet dynamics in a wind-like CSM}
\begin{figure*}
\begin{center}
\includegraphics[scale=2.1,angle=90]{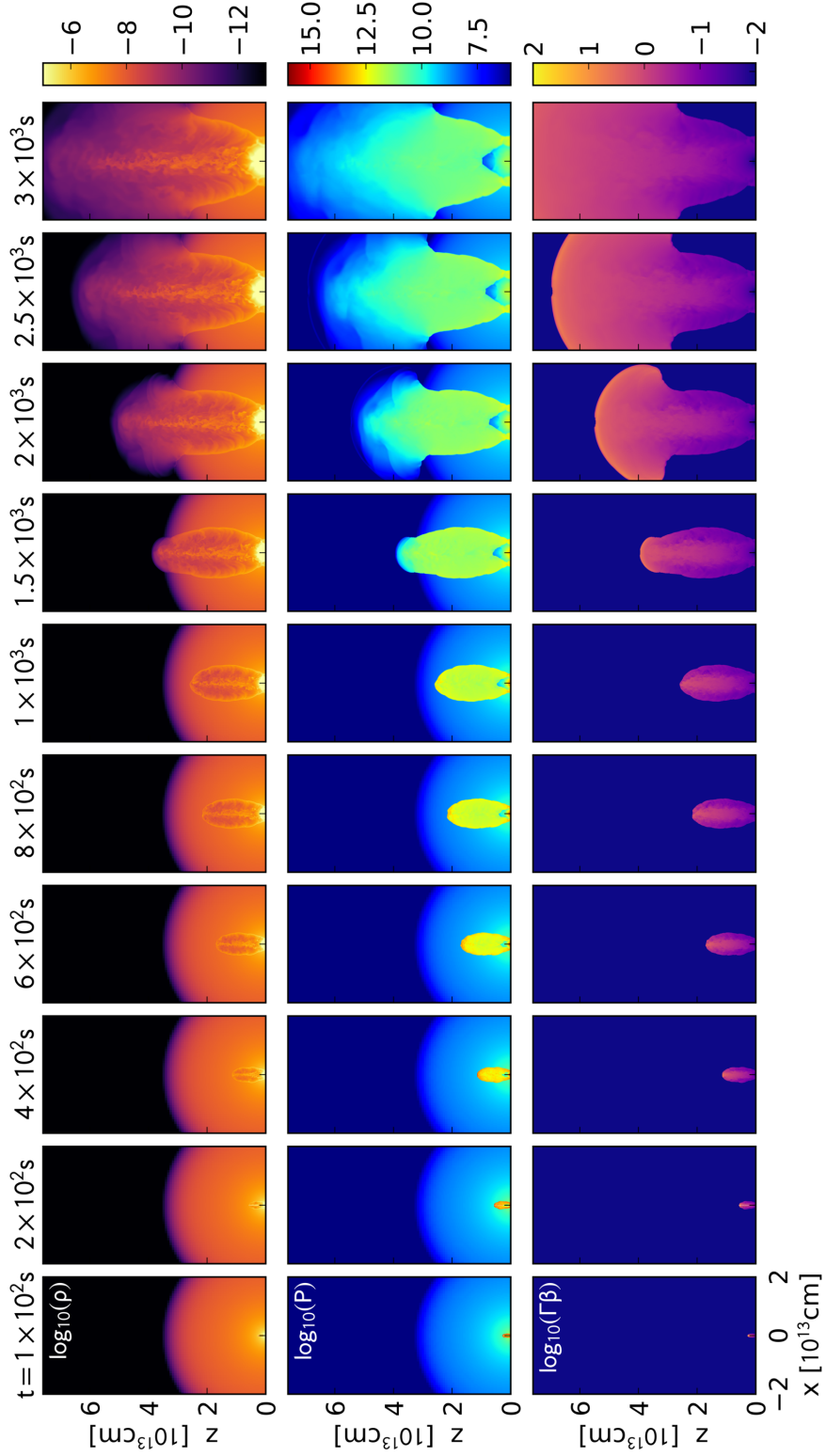}
\end{center}
\caption{Jet propagation in a $1M_\odot$ CSM with $R_\mathrm{csm}=400\,R_\odot$. 
The spatial distributions of the density (top; in $\mathrm{g}\,\mathrm{cm}^{-3}$), the pressure (middle; in $\mathrm{erg}\,\mathrm{cm}^{-3}$), and the 4-velocity (bottom) are compared. 
From left to right, 10 snapshots from $t=10^2$ to $3\times 10^3$ s are presented. 
}
\label{fig:evolution}
\end{figure*}
\begin{figure*}
\begin{center}
\includegraphics[scale=2.1,angle=90]{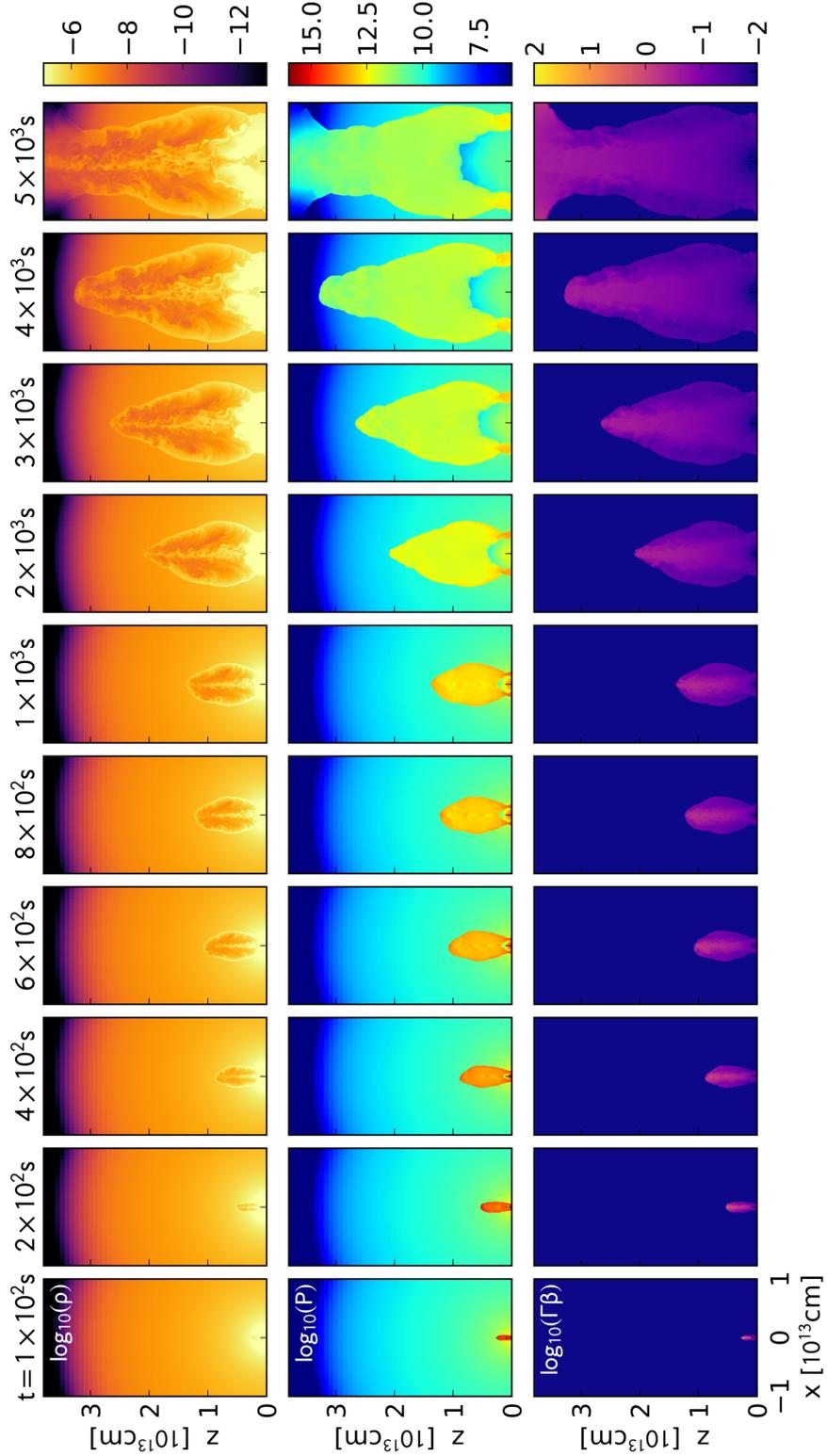}
\end{center}
\caption{Jet propagation in a  $M_\mathrm{csm}=10\,M_\odot$ CSM with $R_\mathrm{csm}=400\,R_\odot$. 
Physical variables are plotted in a similar way to Figure \ref{fig:evolution}, but snapshots with a different scale at different epochs from $t=10^2$ to $5\times 10^3\,\mathrm{s}$ are presented. 
}
\label{fig:evolution_M10R400}
\end{figure*}
\begin{figure*}
\begin{center}
\includegraphics[scale=0.66]{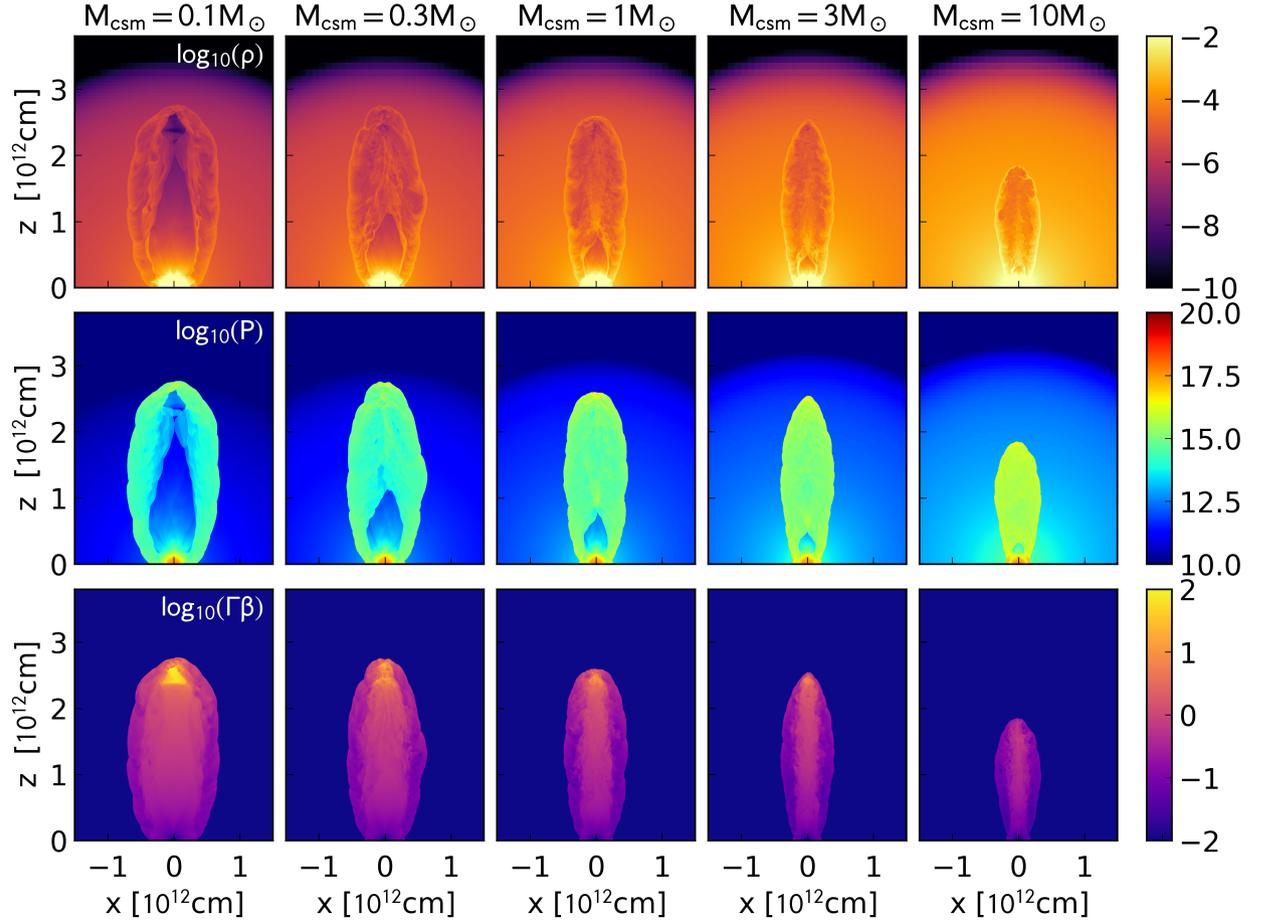}
\end{center}
\caption{Jet morphology in models with different CSM mass. 
The spatial distributions of the density (top), the pressure (middle), and the 4-velocity (bottom) are compared. 
From left to right, the snapshot at $t=10^2\,$s from the models with $M_\mathrm{csm}=0.1,$ $0.3$, $1.0$, $3.0$, and $10\,M_\odot$ and $R_\mathrm{csm}=40\,R_\odot$ are presented. 
}
\label{fig:comparison}
\end{figure*}
\begin{figure*}
\begin{center}
\includegraphics[scale=0.66]{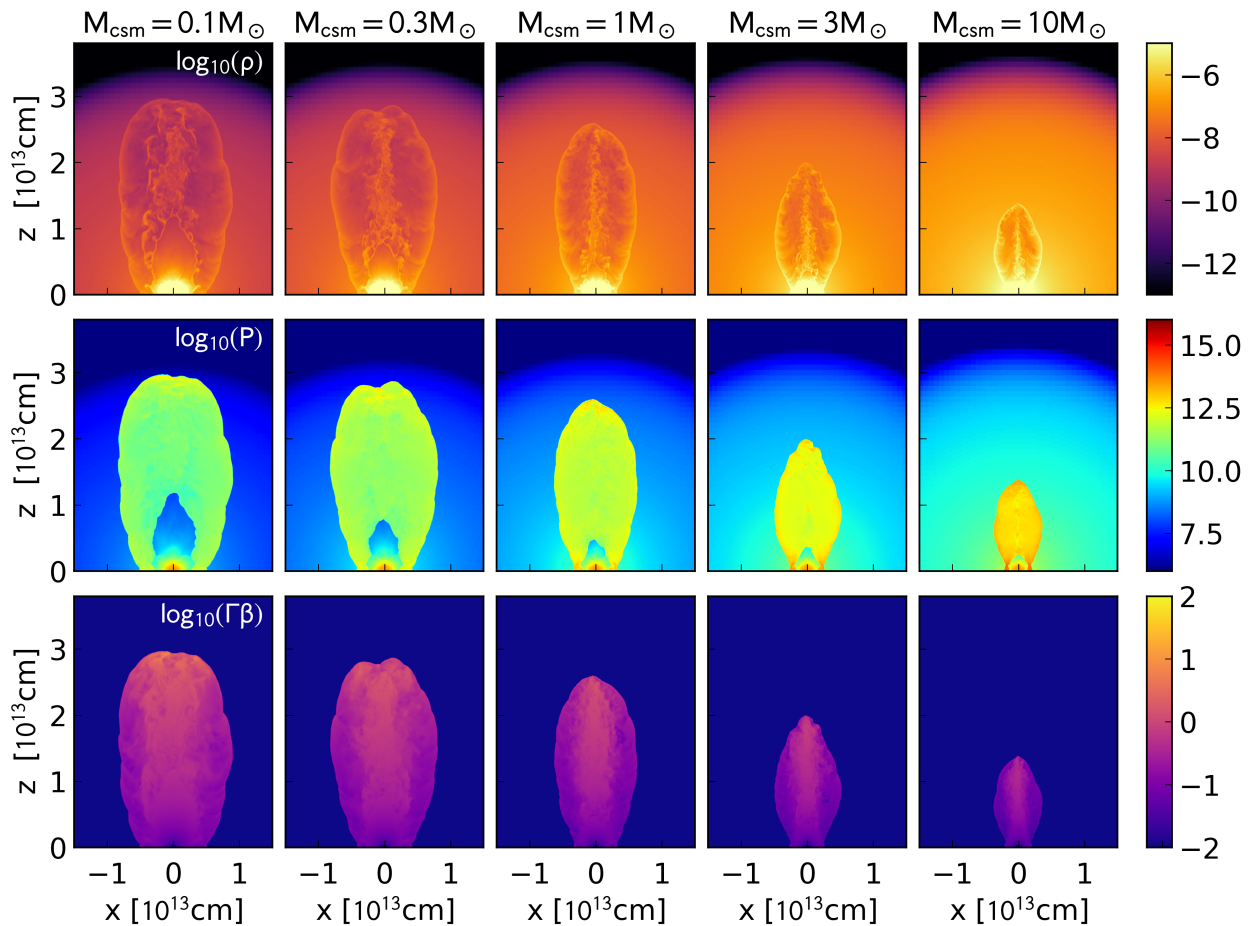}
\end{center}
\caption{Same as Figure \ref{fig:comparison}, but for models with $R_\mathrm{csm}=400\,R_\odot$ at $t=10^3\,\mathrm{s}$. 
}
\label{fig:comparison_R400}
\end{figure*}

Since we adopt the same progenitor model as our previous study, the jet dynamics within the star is same as the standard jet model in \citet{2022ApJ...925..148S}. 
Notable differences are instead found in the jet propagation outside the star. 

Figure \ref{fig:evolution} shows the propagation of the jet in a $1\,M_\odot$ CSM with $R_\mathrm{csm}=400\,R_\odot$ (the \verb|M1R400| model) from $t=10^2\,$s to $3\times 10^3\,$s. 
Without any dense material outside the star, the jet emerging from the stellar surface would be accompanied by the fast expansion of the shocked stellar materials (referred to as the ``stellar cocoon'' hereafter) into the direction normal to the jet axis, making quasi-spherical sub--relativistic ejecta. 
In the presence of the massive CSM, however, the jet and the stellar cocoon are still confined after penetrating the star. 
The high ram pressure of the massive CSM exerting on the jet limits the lateral expansion of the stellar cocoon. 
This leads to a jet highly elongated along the jet axis. 
This is associated with the formation of a secondary cocoon (the ``CSM cocoon''), which is composed of the CSM swept by the forward shock. 
As seen in the pressure distribution in Figure \ref{fig:evolution}, a recollimation shock develops and bends the course of the ejecta toward the $z$-axis. 
The ejected materials are still relativistic, but the 4-velocity distribution shows that it accelerates only up to $\Gamma\beta\simeq 10$ in the CSM. 
The collimated ejecta propagates through the CSM until the forward shock reaches the outer edge of the CSM, at $r=R_\mathrm{csm}$. 
After the shock emergence from the CSM surface, the CSM cocoon starts spreading in the dilute ambient gas in a quasi-spherical way. 
This is understood in an analogy to the stellar cocoon expansion after the jet emergence from a bare star. 
The oblique shock driven by the jet progressively sweeps the outer layers of the CSM at larger inclination angles from the jet axis. 
The ejection of the outer layers immediately creates the negative pressure gradient in the ejecta along the radial direction, which continuously accelerates the ejecta toward the ambient dilute space. 
As such, the energy originally carried by the jet is first dissipated into the thermal energy in the cocoon and then redistributed to the kinetic energy of the ejecta. 
The quasi-spherically expanding ejecta soon enters the free-expansion regime after the internal energy redistribution and then the density structure freezes out. 

A more massive CSM  decelerates the jet to slower velocities more significantly. 
Figure \ref{fig:evolution_M10R400} shows the jet morphology and evolution for the model with $M_\mathrm{csm}=10\,M_\odot$ and $R_\mathrm{csm}=400\,R_\odot$ (the \verb|M10R400| model) from $t=10^2\,\mathrm{s}$ to $5\times 10^3\,\mathrm{s}$. 
This is a representative model for a jet with heavy mass-loading and deceleration to sub-relativistic velocities. 
The jet height almost linearly grows up to $t\simeq 200\,\mathrm{s}$ with the forward shock speed being close to the speed of light. 
Then, the forward shock (along the jet axis) slows down to non-relativistic speeds. 
Even at $t\simeq 10^3\,\mathrm{s}$, the forward shock front is still located at $z\simeq 1.3\times 10^{13}\,\mathrm{cm}$ (i.e., $z/t\sim 0.4c$). 
The forward shock reaches the CSM outer surface at $\simeq 5\times 10^3\,\mathrm{s}$, significantly later than $R_\mathrm{csm}/c\simeq 10^3\,\mathrm{s}$. 
We shall discuss the jet evolution in more detail in Sec. \ref{sec:jet_analysis}. 

Lastly, we remark on the numerical resolution and its influence on the simulation results. 
Our simulations employ the AMR technique, which appears to successfully resolve small structures produced in simulations as seen in Figures \ref{fig:evolution} and \ref{fig:evolution_M10R400}. 
Nevertheless, hydrodynamic instabilities, such as the Kelvin-Helmholtz instability developing at the interface between the jet and the ambient medium, could produce structures even smaller than the minimum resolved length. 
This influences the conversion of the jet kinetic energy into the cocoon thermal pressure and the mixing of the jet, stellar, and CSM materials, leading to quantitative differences between simulations with the same setup but different numerical resolutions. 
For example, higher cocoon pressure makes the jet more collimated and keeps the jet head faster. 
The relativistic ejecta forms as a result of the forward shock passage followed by the acceleration and expansion of the CSM outer layers. 
Therefore, while the resultant ejecta structure would be insensitive to the shock velocity, the amount of mass and energy loaded on the ejected material can be dependent on the shock velocity. 
We also note that, in our previous work (\cite{2022ApJ...925..148S}), we tried models with lower resolution and find no qualitative difference (though not explicitly presented).

\subsubsection{Dependence on CSM mass} 
The jet morphology is highly dependent on the CSM mass. 
In Figures \ref{fig:comparison} and \ref{fig:comparison_R400}, we compare the jet morphology for models with the same CSM radius ($R_\mathrm{csm}=40\,R_\odot$ and $400\,R_\odot$, respectively) but with different CSM masses at the same epoch. 
After the launch of the highly relativistic materials from the stellar surface, the jet slows down to lower velocities due to the ambient ram pressure.  
For a fixed CSM radius, an increased CSM mass leads to a larger ambient density and thus more efficiently prevents the expansion of the ejected material. 
This difference makes the distributions in Figure \ref{fig:comparison} progressively more complex for less massive CSMs (panels on the left) than for more massive CSMs (panels on the right). 

For large CSM masses, the jet is tightly collimated. 
The discontinuous pressure distribution in each column in 
Figures \ref{fig:comparison} and \ref{fig:comparison_R400} indicates the location of the recollimation shock. 
Models with larger CSM masses show a smaller extent in the collimation shock  compared to the cocoon, due to the deeper penetration of the shock; it makes almost all of the ejecta shocked for the most massive CSM cases. 
The shocked region is predominantly occupied by the shocked CSM, which was initially smoothly distributed. 
In addition, once the ejecta and CSM are swept by the shock, the post-shock gas tries to achieve pressure balance. 
This leaves the density and pressure distributions of massive CSM models relatively featureless. 

Models with smaller CSM masses show more complicated spatial structures. 
For the smallest CSM mass $M_\mathrm{csm}=0.1M_\odot$ and $R_\mathrm{csm}=40\,R_\odot$ (\verb|M01R40| model), the jet  
still remains highly relativistic with $\Gamma\beta>10$ at $t=10^2$ s (the leftmost column in Figure \ref{fig:comparison}). 
Therefore, the jet height is close to $z=c(t-t_\mathrm{br})$ with the jet breakout time {(from the stellar surface) being $t_\mathrm{br}\simeq 5$ s. 
In this model, a part of the jet has not yet been swept by the reverse shock, as evidenced by the high velocity spot (jet tail) in Figure \ref{fig:comparison}. 
We focus on the issue of shock propagation and its dependence on the CSM mass in more detail in Section \ref{sec:jet_analysis}.

The ejecta deceleration is most effective for laterally expanding ejecta carrying smaller kinetic energy per a unit solid angle than those around the jet axis.  
We indeed observe narrower shocked regions for more massive CSMs. 
This trend indicates that a smaller fraction of the CSM is swept by the shock for a more massive CSM with the same radius. 
Therefore we expect the mass-loading of the jet would not simply scale with the CSM mass in a linear fashion. 
In fact, the mass of the materials accelerated beyond $\Gamma\beta=0.1$ is only weakly dependent on the CSM mass, as we shall see below. 

The recent 2D simulations of GRB jets in dense environments by \citet{2020ApJ...900..193D} have revealed efficient jet deceleration for their models with $3\,M_\odot$ and $30\,M_\odot$ CSM within $10^{13}\,\mathrm{cm}$. 
They suggest that the jet deceleration is significant for the CSM density coefficient of $A_\mathrm{csm}>4\times 10^{19}\,\mathrm{g}\,\mathrm{cm}^{-1}$. 
Our simulations with a similar CSM radius of $400\,R_\odot\simeq 3\times 10^{13}\,\mathrm{cm}$ show that jets in $3\,M_\odot$ and $10\,M_\odot$ CSM significantly decelerate down to non-relativistic velocities  (Figure \ref{fig:comparison_R400}), which agrees with the simulation results of \citet{2020ApJ...900..193D}.

\subsubsection{Averaged radial profiles}\label{sec:averaged_radial_profiles}
\begin{figure*}
\begin{center}
\includegraphics[scale=0.7]{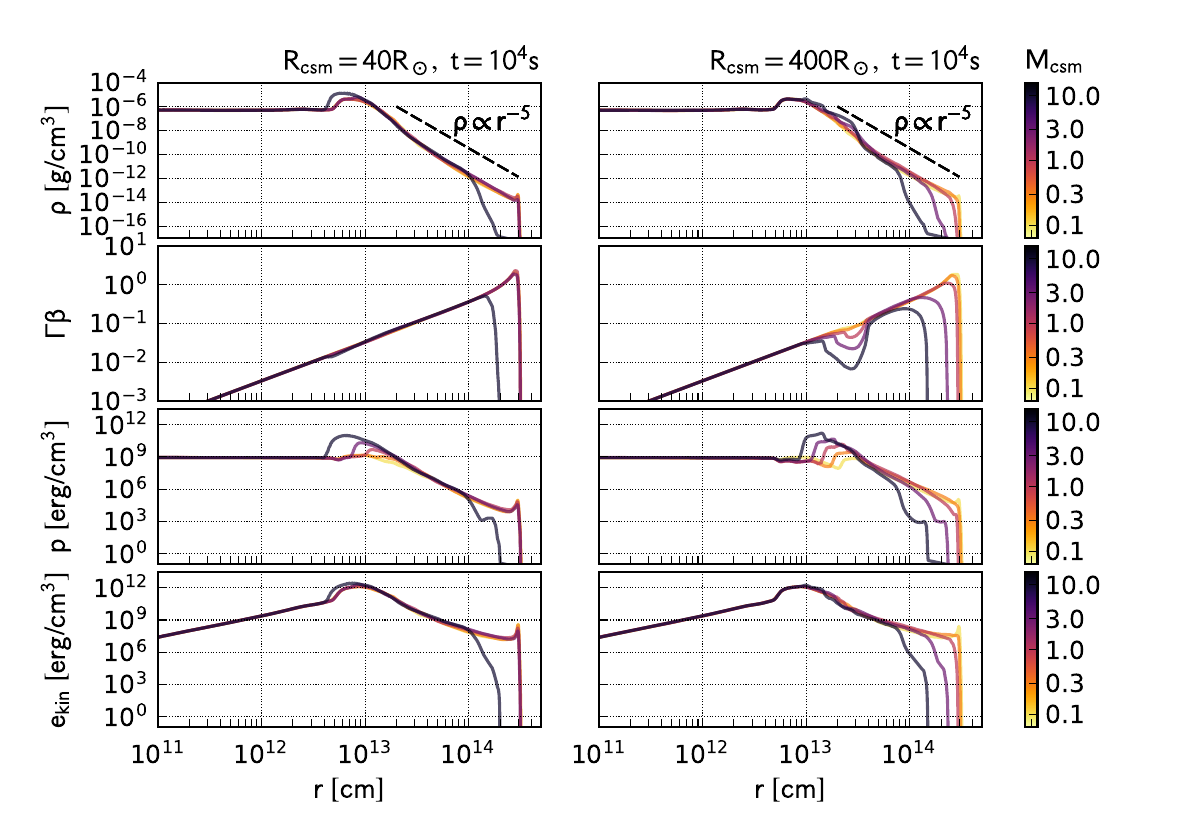}
\end{center}
\caption{Angle-averaged radial profiles for the models with $R=40\,R_\odot$ (left) and $400\,R_\odot$ (right). 
The density, four-velocity, and pressure profiles are plotted from top to bottom. 
The profiles in each panel are colour-coded in terms of the CSM mass. 
In the top panels, a power-law density profile with an index of $-5$ is plotted for comparison. }
\label{fig:radial}
\end{figure*}

Next, we investigate the dynamical properties of the mildly relativistic ejecta well after the shock emergence from the CSM surface. 
Figure \ref{fig:radial} shows the angle-averaged radial profiles of several hydrodynamic variables at $t=10^4\,\mathrm{s}$. 
The angle-averaged value of a variable $q(x,y,z)$ is computed in the following integration over $2\pi$ solid angle,
\begin{equation}
    \langle q\rangle=\frac{1}{2\pi}\int^1_0\int_0^{2\pi} q(x,y,z)\mathrm{d}\cos\theta \mathrm{d}\phi.
\end{equation}
The bottom panels of Figure \ref{fig:radial} presents the angle-averaged kinetic energy density $\langle e_\mathrm{kin}\rangle$, with $e_\mathrm{kin}$ defined as
\begin{equation}
    e_\mathrm{kin}=\rho\Gamma(\Gamma-1)c^2. 
\end{equation}

For models with $R_\mathrm{csm}=40\,R_\odot$ (the left column of Figure \ref{fig:radial}), the radial profiles at larger radii show remarkable similarities among models with different amounts of CSM. 
Except for the most massive $M_\mathrm{csm}(=10M_\odot)$, the velocity profile extends beyond $\Gamma\beta>1$. 
The density profiles exhibit inner flat and outer steep parts. 
We note that a part of the ejecta exhibits even higher Lorentz factor around the jet axis. 
Such high $\Gamma$ region is, however, located only around the jet axis and thus does not predominantly contribute to the radial profiles after the angle average. 
The model with $M_\mathrm{csm}=10M_\odot$ is the only model ending up with a sub-relativistic jet head for $R_\mathrm{csm}=40\,R_\odot$, where
the 4-velocity profile is truncated around $\Gamma\beta\simeq 0.5$. 
Even with different maximum 4-velocities, the density profiles for $\Gamma\beta>0.1$ are well represented by a single power-law function with $\mathrm{d}\ln\rho/\mathrm{d}\ln r\simeq -5$. 
Assuming free expansion $r=vt$, this is equivalent to $\rho\propto v^{-5}$. 
This characteristic radial structure is also seen in our previous models (the standard and choked jet models in \citet{2022ApJ...925..148S}). 
Similar conclusions have been reached in the recent work by \citet{2023MNRAS.519.1941P}, who conducted a series of 2D cylindrical GRB jet simulations with various settings. 

In a non-relativistic freely expanding regime, this power-law density profile $\rho\propto v^{-5}$ is equivalent to a flat kinetic energy distribution $\mathrm{d} E/\mathrm{d}\ln v\propto \rho v^{5}\propto$ Const., i.e., the same amount of energy is distributed in equidistant shells in the logarithmic velocity space. 
Recently, \citet{2022MNRAS.517..582E} conducted 2D GRB jet simulations and demonstrated that flat kinetic energy distributions $\mathrm{d} E/\mathrm{d}\ln(\Gamma\beta)$ are realized in an even higher 4-velocity range up to $\Gamma\beta\simeq 3$. 
These agreements in GRB jet simulations with different simulation codes and settings suggest that mildly relativistic ejecta with a density slope of $\mathrm{d}\ln\rho/\mathrm{d}\ln v=-5$ may be ubiquitously realized in association with the jet emergence from a slowly moving or static dense material. 
This is also perhaps true in a broader class of SN explosions harboring a long-lasting engine with a constant power \citep{2017MNRAS.466.2633S,2019ApJ...880..150S}. 

The models with a more extended CSM ($R_\mathrm{csm}=400\,R_\odot$) also exhibit similarly universal density structure but with more diverse high-velocity cutoffs (the right column in Figure \ref{fig:radial}). 
The two models with $M_\mathrm{csm}=3\,M_\odot$ and $10\,M_\odot$ show maximum 4-velocities of $\simeq 0.5c$ and $0.2c$, respectively. 
A more extended CSM continues to exert high ram pressure on the ejecta for a longer period and thus more efficiently decelerates the jet. 
Each 4-velocity distribution has a dip at the assumed outer CSM radius. 
This is because the ejecta is still redistributing its kinetic energy at $t=10^{4}$ s, as the free expansion has not yet been completely established for these models with the extended CSM. 
After the jet emergence from the CSM, the ejected materials expand in a quasi-spherical way and start covering the CSM, and these ejecta constitute the outer freely expanding material beyond the dip.  
On the other hand, the initial thermal bomb and a fraction of the jet energy, which have been dissipated in the star and CSM, make the stellar materials accelerate. 
This constitutes the inner freely expanding material within the dip. 
After sufficiently long time for relaxation, the injected energy is shared by the whole ejecta and the entire velocity distribution follows $r/t$. 
The non-relativistic jets in these two models take longer times to reach the outer CSM radius, which significantly delays the completion of the energy redistribution and the freeze-out of the density structure. 
Nevertheless, the outermost layers of each model appear to follow a power-law profile with a slope similar to $\rho\propto r^{-5}$. 
This indicates that the density profile for this high velocity ejecta has already frozen out. 
For the models with $R_\mathrm{csm}=400\,R_\odot$ (right panels), the pressure and kinetic energy density in Figure \ref{fig:radial} are indeed comparable with each other around $r=2$--$3\times 10^{13}\,\mathrm{cm}$, while the pressure is negligible compared with the kinetic energy density at outer layers with $r>3\times 10^{13}\,\mathrm{cm}$. 

\begin{figure*}
\begin{center}
\includegraphics[scale=0.7]{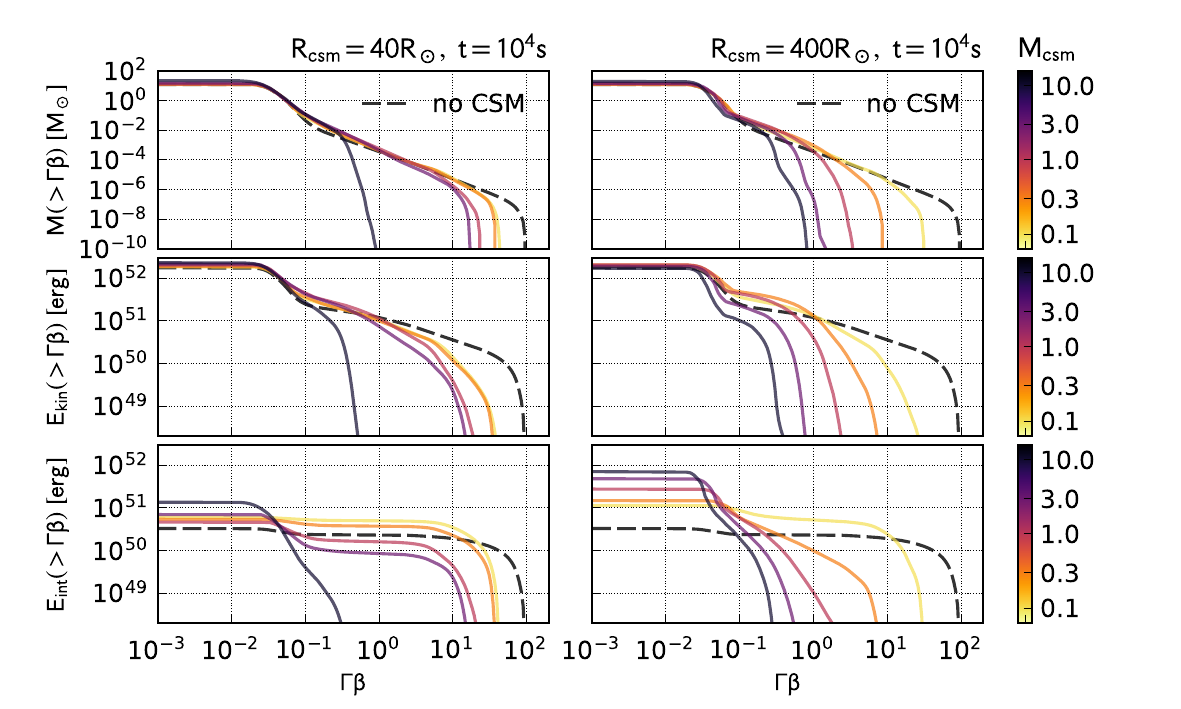}
\end{center}
\caption{Mass and energy spectra for the models with $R=40\,R_\odot$ (left) and $400\,R_\odot$ (right). 
The mass, kinetic energy, and internal energy distributions are plotted from top to bottom. 
The distributions in each panel are colour-coded in terms of the CSM mass. 
The black dashed curves represent the no CSM model presented in \protect\citet{2022ApJ...925..148S}
.}
\label{fig:spec}
\end{figure*}
\subsubsection{Mass and energy spectra}
As in the previous work, we calculate the mass, the kinetic energy, and the internal energy distributions of the ejecta traveling at 4-velocities faster than a threshold value $\Gamma\beta$,
\begin{equation}
    M(\Gamma\beta)=\int_{>\Gamma\beta}\rho\Gamma \mathrm{d}V,
\end{equation}
\begin{equation}
    E_\mathrm{kin}
    (\Gamma\beta)=
    \int_{>\Gamma\beta}\rho\Gamma(\Gamma-1) \mathrm{d}V,
\end{equation}
and
\begin{equation}
    E_\mathrm{int}(\Gamma\beta)=\int_{>\Gamma\beta}
    \left(\frac{\gamma}{\gamma-1}\Gamma^2-1\right)p\mathrm{d}V.
\end{equation}
Here, the volume integration is carried out only for numerical cells with the 4-velocity larger than a given value of $\Gamma\beta$. 
Figure \ref{fig:spec} shows the mass and energy distributions for all the models. 
The distributions are compared with the model without a massive CSM in \citet{2022ApJ...925..148S} (black dashed lines). 

As was pointed out by \citet{2022ApJ...925..148S}, the mass distributions show a flat non-relativistic part and a relativistic power-law part with a high-velocity cutoff. 
These two segments are connected around $\Gamma\beta\simeq 0.1$. 
In contrast to the no CSM model extending to the maximum 4-velocity of $\Gamma\beta\sim 100$, the models in this work show progressively lower maximum 4-velocities for larger CSM masses. 

Despite the clear difference in the velocity cut-offs, the mass and energy distributions in the non- and mildly relativistic regimes are similar to each other. 
The ejecta mass with a velocity exceeding $0.1c$ is more or less $0.1\,M_\odot$ for different CSM masses. 
The corresponding kinetic energies are of the order of $\simeq 10^{51}\,\mathrm{erg}$. 
In Figure \ref{fig:csm_dependence}, we plot the mass and the kinetic energy of the ejecta with a 4-velocity faster than $0.1c$ as a function of the CSM mass. 
The mass $M(\Gamma\beta>0.1)$ only slightly increases as a function of the CSM mass for $M_\mathrm{csm}=0.1\,M_\odot$--$1\,M_\odot$ and then stays constant ($R_\mathrm{csm}=40\,R_\odot$) or slightly decreases ($R_\mathrm{csm}=400\,R_\odot$) for $M_\mathrm{csm}=3$ and $10\,M_\odot$. 
The corresponding kinetic energy is $(4$--$6)\times 10^{51}\,\mathrm{erg}$ for $M_\mathrm{csm}=0.1$--$1M_\odot$, while it decreases to smaller values for  $M_\mathrm{csm}=3$ and $10\,M_\odot$. 
The decrease in the mass and energy for higher CSM masses indicates the deceleration of the ejecta to sub-relativistic velocities. 
Overall, however, the high-velocity ejecta with $\Gamma \beta>0.1$ shows similar masses and kinetic energies, despite the large difference in the CSM mass covering two orders of magnitude within the parameter range explored. 

We note that an even more massive CSM/envelope ($>10M_\odot$) would more efficiently decelerate the jet and thus accelerate a smaller amount of ejecta to sub-relativistic velocities, but locating more than tens of solar masses (i.e., more than the maximum CSM mass in our simulations) within blue or red supergiant radii might require extreme conditions. 


\subsection{Jet morphology and choking}\label{sec:jet_analysis}
The propagation of a relativistic jet in ambient gas and the cocoon formation as a result of the jet-ambient gas interaction have been extensively studied in analytic ways (e.g., \cite{1989ApJ...345L..21B,1997ApJ...479..151M,2003MNRAS.345..575M,2011ApJ...740..100B,2013ApJ...777..162M,2019MNRAS.489.2844I,2020ApJ...900..193D}).  
As has been investigated by these previous works, the forward shock along the jet axis is initially driven by the continuously injected jet (the ram pressure or the momentum), while the lateral expansion is governed by the thermal pressure of the shocked gas (the stellar cocoon). 
After the jet injection is terminated, however, the jet head continues to slow down, eventually making the entire shock driven by the thermal pressure. 
\citet{2019MNRAS.489.2844I} assumed that the outflow is predominantly driven by its thermal pressure and investigated the shock evolution in power-law media. 
In their results, the shocked gas initially elongated along the jet axis gradually approaches a spherical Sedov-Taylor solution for a steady wind ($\rho\propto r^{-2}$) or a shallower power-law radial density profile. 
Our simulations of jet-CSM interaction cover transitional cases toward such a pressure-driven shock. 

As the jet morphology in Figures \ref{fig:comparison} and \ref{fig:comparison_R400} suggests, the jets are well confined in the massive CSM for the explored CSM parameters (i.e., the collimated jet regime in \citet{2011ApJ...740..100B}). 
Then, the jet propagation can further be divided into relativistic and non-relativistic regimes. 
For a less massive CSM, the forward shock is still relativistic at the emergence from the CSM outer radius. 
As long as the jet is in the relativistic regime, the jet height is given by $z\simeq c(t-t_\mathrm{br})$. 
An increase in the CSM mass results in a narrower CSM cocoon due to the higher ambient ram pressure. 
This is highlighted by the models with $M_\mathrm{csm}=0.1$--$1\,M_\odot$ in Figure \ref{fig:comparison} (three left columns). 
For jets in the relativistic regime, the width at the same epoch roughly halves for an increase in the CSM mass by a factor of $10$. 

In an even more massive CSM, the forward shock along the jet axis slows down to non-relativistic velocities at some time before reaching the CSM outer radius. 
Figure \ref{fig:evolution_M10R400} represents the jet evolution in the model with the most massive and extended CSM, which realizes a non-relativistic forward shock. 
In the deceleration stage, the initially narrow shocked region widens due to the lateral expansion. 
The height reaches $z\simeq 1.3\times 10^{13}\,\mathrm{cm}$ at $t=10^3\,\mathrm{s}$, i.e., it is only doubled compared to the height at $t=2\times 10^2\,\mathrm{s}$. 

This transition from relativistic to non-relativistic shock velocities is accompanied by changes in the shape of the shocked region and its pressure distribution. 
Initially, the overall shape of the shocked region is elongated, and the pressure distribution is non-uniform, with significantly higher pressure and shock velocity near the axis.  
Even after all of the jet tail is swept by the reverse shock, the high on-axis pressure and velocity persist temporarily, so that the outflow remains relatively elongated.
The pressure distribution then gradually becomes uniform as the lateral expansion of the shocked gas catches up with the shock along the jet axis. 
The models with non-relativistic shock velocity along the jet axis (e.g., the rightmost two models in Figure \ref{fig:comparison_R400}) show relatively uniform pressure distributions in the shocked gas. 

\citet{2019MNRAS.489.2844I} suggest that once the outflow's width and height become comparable, the shock then enters another expansion stage, with its time-dependence given by that of a Sedov-Taylor blast wave. 
At $t>1\times 10^3\,\mathrm{s}$, the jet shape indeed makes a further change. 
Although the shock around the jet axis is still elongated due to the steep density gradient, most parts of the shock front appear to be predominantly driven by the thermal pressure of the shocked gas. 
In the Sedov-Taylor blast wave regime, the forward shock radius should evolve as
\begin{equation}
    R_\mathrm{fs}\sim
    \left(\frac{E_\mathrm{jet}t^2}{A_\mathrm{csm}}\right)^{1/3}\simeq 
    \left(\frac{4\pi E_\mathrm{jet}R_\mathrm{csm}t^2}{M_\mathrm{csm}}\right)^{1/3}
    ,
\end{equation}
for $\rho=A_\mathrm{csm}r^{-2}$, from dimensional analysis. 
This scaling appears to hold true for the lateral parts of the shock in Figure \ref{fig:evolution_M10R400}. 
With the time dependence $R_\mathrm{fs}\propto t^{2/3}$, we expect an increase in the shock radius by a factor of $2^{2/3}\simeq 1.6$ from $t=2\times 10^{3}\,\mathrm{s}$ to $4\times 10^{3}\,\mathrm{s}$. 
This is indeed consistent with the expansion rate of the shocked region in Figure \ref{fig:evolution_M10R400} during the corresponding period. 
The dependence on the CSM mass, $R_\mathrm{fs}\propto M_\mathrm{csm}^{-1/3}$, suggests that the jet height and width are smaller by a factor of $10^{-1/3}\simeq 2.15$ for an increase in the CSM mass from $1\,M_\odot$ to $10\,M_\odot$. 
This is consistent with the shocked region in the models with $M_\mathrm{csm}=1$--$10\,M_\odot$ in Figure \ref{fig:comparison_R400}. 
Therefore, the dependence on the jet morphology on time and the ambient density would be approximately given by those of the Sedov-Taylor self-similar solution in this regime.

In summary, the mass-loading of a GRB jet in a massive CSM proceeds in the following evolutionary stages (see, Figure \ref{fig:jet_schematic}): 
(1) Soon after the jet emergence from the stellar surface, the jet is uncollimated and freely expands with a Lorentz factor determined by the injection condition. 
(2) The high ram pressure of the CSM leads to the formation of the CSM cocoon and the recollimation shock, confining ejected materials into a narrow region around the jet axis. 
A more massive CSM forms a smaller recollimation shock and confines the shocked materials more tightly. 
(3)  After the central engine shuts off,  the reverse shock formed as a result of the jet-CSM collision still propagates along the jet axis. 
The bottom part of the jet (jet tail) is not shocked yet. 
The jet head continues to be driven forward by the ram pressure supplied by the jet tail, resulting in a high-pressure region near the jet axis. 
(4) Even after the reverse shock sweeps the entire jet tail, the outflow still temporarily keeps its elongated shape as long as the high pressure region near the jet axis persists. 
As the internal energy near the axis is redistributed throughout the entire shocked gas, the forward shock along the axis decelerates. 
(5) Eventually, the pressure in the outflow becomes uniform and the shock is driven by thermal pressure even along the axis.  Once the on-axis and lateral shock velocities become comparable, the outflow starts transitioning to a spherical flow. 
After an initial sideways expansion phase, the shock front and the shocked gas become quasi-spherical and 
then evolve in a self-similar fashion. 

The time-scale of each evolutionary stage is determined by the jet and CSM properties. 
Stage (2), in which the jet is still being injected, ends when the jet injection is terminated ($t_\mathrm{in}=20\,\mathrm{s}$ in the simulations). 
The transition between stages (3) and (4) happens when the jet tail catches up with the jet head, which occurs at a time $t_\mathrm{c} = \beta_\mathrm{t} t_\mathrm{in}/(\beta_t-\beta_\mathrm{h}) \approx 2 \Gamma_\mathrm{h}^2 t_\mathrm{in}$ (e.g., \cite{2015ApJ...807..172N}), where $\beta_\mathrm{h}$ and $\Gamma_\mathrm{h}$ are the velocity and Lorentz factor of the jet head, and $\beta_\mathrm{t} \approx 1$ is the velocity of the jet tail. 
Therefore, the relative velocity of the jet tail
and the jet head, which is dependent on the jet power and the CSM density (or $M_\mathrm{csm}$ and $R_\mathrm{csm}$), determines the end of stage (3). 
Then, the jet makes a transition from  stage (4) to stage (5) when the pressure imbalance is resolved. 
In order to achieve uniform pressure in the shocked gas, the information at the forward shock along the axis should propagate back through the expanding flow. 
The redistribution of the internal energy presumably 
requires several sound crossing times across the shocked region. 
Thus, the time when the transition between stages (4) and (5) occurs is more ambiguous compared to the transitions between the other stages. 
Therefore, further analytic and numerical investigations are encouraged.

We note that the jet propagation in a CSM proceeds according to the evolutionary stages above, as long as the dense CSM with a shallow density profile extends infinitely. 
In reality, however, the dense and shallow CSM would be truncated somewhere and the ambient density would drop down to a small value, like the normal stellar wind as is assumed in our simulations. 
At the transition layer between the dense and dilute ambient media (corresponding to $r=R_\mathrm{csm}$ in our simulations), the forward shock accelerates and the dynamical evolution would be different from the above evolutionary stages.

Our simulations assume a massive enough CSM for the jet to immediately enter the evolutionary stages (2)--(5) after stage (1). 
With no or only a tiny CSM, the ejected materials would end up in stage (1) with little effect of the jet-CSM interaction. 
In a compact and/or less massive CSM, a jet may reach the outer radius of the CSM before entering stages (4) and (5). 
We find that most cases end up in the intermediate evolutionary stage (4) for the explored CSM parameters. 
The model with the least massive and extended CSM (model \verb|M01R40|; the leftmost panels in Figure \ref{fig:comparison}), for example, ends up in stage (3) with an unshocked jet tail. 
Several models with massive and extended CSM (models \verb|M3R400| and \verb|M10R400|; the rightmost two columns in Figure \ref{fig:comparison_R400}) reach stage (5). 
We note that the intermediate stages (3) and (4) are probably more important in cases where the jet head is relativistic. 
Once the forward shock along the jet axis slows down to non-relativistic velocities, the lateral shock, which is non-relativistic in all our simulations, soon catches up with the expansion, leading to stage (5). 
As discussed above, stage (3) begins when $t=t_\mathrm{in}$, and ends when $t=t_\mathrm{c}$, so the duration of stage (3) is given by $t_\mathrm{c} - t_\mathrm{in} \approx 2 \beta_\mathrm{h}\Gamma_\mathrm{h}^2 t_\mathrm{in}$.  Therefore we expect a jet with a relativistic head, $\Gamma_\mathrm{h} \gg 1$, to spend a significant time (much longer than $t_\mathrm{in}$) in stages (3) and (4).  On the other hand, a non-relativistic jet head with $\beta_\mathrm{h} \ll 1$ is expected to spend only a short time in these stages, and quickly evolve to the quasi-spherical phase (5).  This is consistent with our simulation results. 

\begin{figure}
\begin{center}
\includegraphics[scale=0.7]{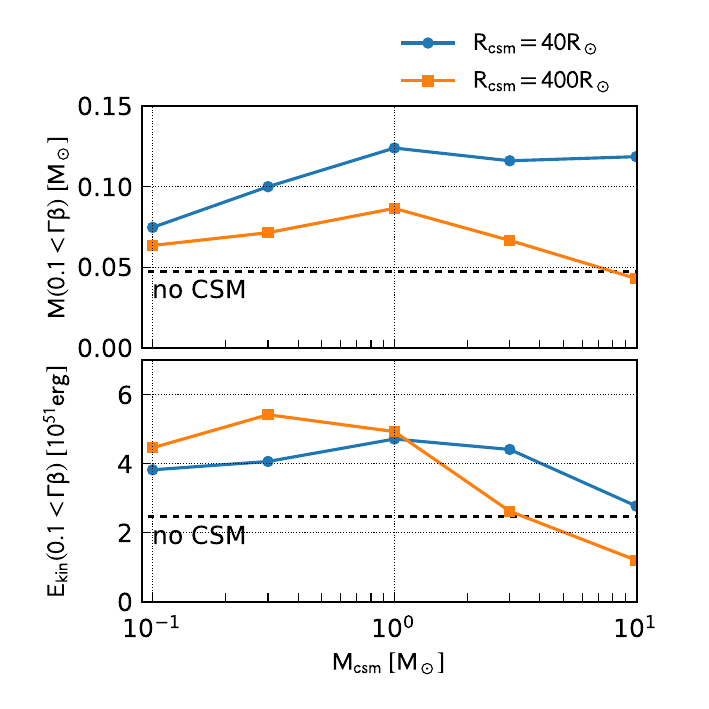}
\end{center}
\caption{CSM dependence on the ejecta accelerated to 4-velocities faster than $0.1c$. 
The mass and the kinetic energy of the ejecta are plotted as a function of the assumed CSM mass in upper and lower panels, respectively. 
In each panel, we present results for different CSM radii (blue circle; $40\,R_\odot$ and orange square; $400\,R_\odot$). 
We also plot the model without massive CSM from \protect\citet{2022ApJ...925..148S}. 
}
\label{fig:csm_dependence}
\end{figure}

\begin{figure}
\begin{center}
\includegraphics[scale=0.4]{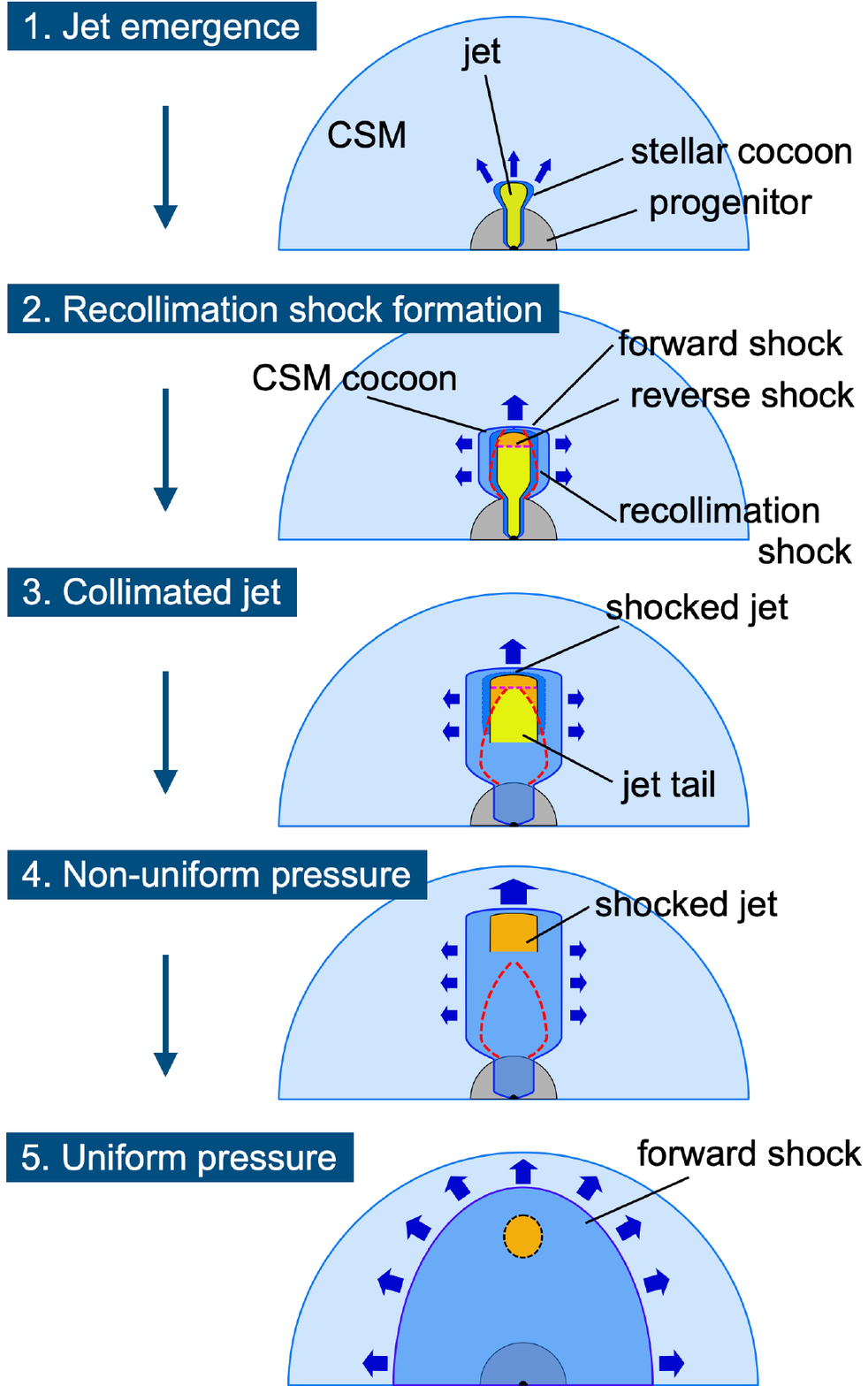}
\end{center}
\caption{Schematic representation of 5 evolutionary stages of a GRB jet propagating in a massive CSM. }
\label{fig:jet_schematic}
\end{figure}

\begin{figure}
\begin{center}
\includegraphics[scale=0.7]{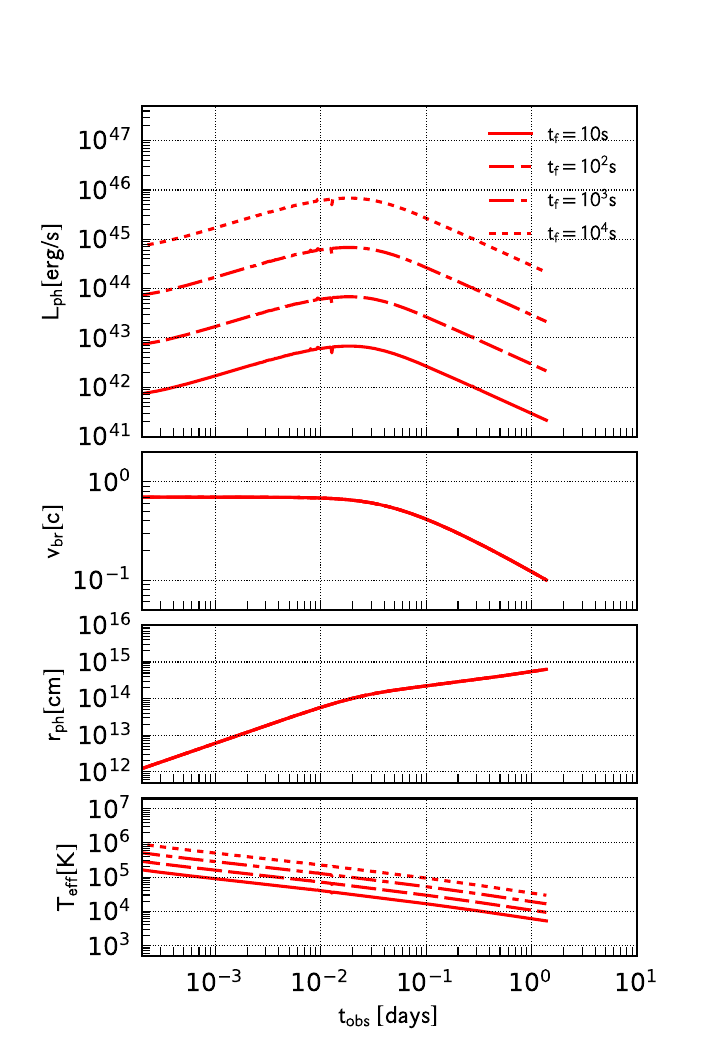}
\end{center}
\caption{Photospheric emission from expanding ejecta with power-law density structure. 
The temporal evolution of the luminosity, the velocity at the breakout layer, the radius of the photosphere, and the effective temperature are plotted from top to bottom. 
The horizontal axis shows the observed time.
}
\label{fig:emission}
\end{figure}

\section{Observational signatures of jet-CSM interaction}\label{sec:radiation}
In this section, we discuss the properties of the relativistic ejecta from jet-CSM interaction from the viewpoint of expected electromagnetic wave signals. 
Our hydrodynamic simulations suggest that $\simeq0.05$--$0.12M_\odot$ of materials are typically accelerated to sub-relativistic speeds ($\Gamma\beta>0.1$) and then ejected with kinetic energies of $\simeq1$--$5\times10^{51}\,\mathrm{erg}$. 
Our simulation results suggest that in the parameter range we explore, 
the mass and kinetic energy of the fast ejecta are not sensitive to the CSM mass and radius (Figure \ref{fig:csm_dependence}), but are probably dependent on the jet properties. 
The power-law density profile, $\rho\propto v^{-5}$, is likely independent of the jet and CSM properties. 
This profile could be widely applied to jet-powered transients, while the density normalization and thus the total mass and energy would depend on the jet properties.  
These findings motivate us to construct emission models assuming fixed density and velocity profiles, but with adjustable normalization constants. 
As we have discussed in \citet{2022ApJ...925..148S} and \citet{2023MNRAS.522.2267M}, this ejecta component commonly manifests itself in some fast-evolving and/or energetic astronomical transients likely involving relativistic jets. 

\subsection{Thermal emission from cooling ejecta}
Thermal photons emitted from the photosphere are expected to provide dominant contribution to early bright emission. 
In the following, we consider the ejecta as spherical and freely expanding gas for simplicity. 
As the ejecta expands, radial layers become dilute and transparent to thermal photons one after another. 
This receding photosphere eventually releases the thermal photons kept in each radial layer. 
The released thermal photons almost freely travel into the interstellar space and are seen as early photospheric emission. 
The following consideration does not include additional heat sources, such as $^{56}$Ni and $^{56}$Co decay. 
Therefore, our model is applicable only in the initial phase up to a few days.

The brightness of the photospheric emission is  dependent on how the photosphere recedes in the ejecta and how much radiation energy is left within the ejecta. 
While the ejecta expands freely, the radiation energy in the optically thick part of the ejecta continues to decrease due to adiabatic loss. 
Assuming that ejecta is created at various formation radii with a fixed amount of initial radiation energy and becomes transparent at a similar radius, the expected emission becomes more (less) luminous when the ejecta has expanded from a large (small) formation radius. 

Thus, the dynamical evolution and the opacity in the ejecta determine the brightness of the thermal emission. 
The thermal emission from cooling ejecta has been observed in some SNe in their infant stages and thus theoretical models have also been investigated extensively (e.g., \cite{1977ApJS...33..515F,1992ApJ...393..742E,2010ApJ...725..904N}). 
In this work, we present a model specialized for our high-velocity, power-law ejecta (see also, \cite{2019ApJ...870...38S}). 

We describe the details of the light curve model in Appendix \ref{sec:light_curve}. 
We consider freely expanding spherical ejecta with a power-law density profile, Equation \ref{eq:density}. 
The ejecta extends from the minimum velocity $\beta_\mathrm{min}$ to the maximum velocity $\beta_\mathrm{max}$, which are fixed to be $(\beta_\mathrm{min},\beta_\mathrm{max})=(0.1,0.7)$.  
In the following, we fix the density slope to be $n\equiv-\mathrm{d}\ln\rho/\mathrm{d}\ln r=5$. 
The radiation energy distribution is also assumed to be a power-law function of radius, Equation \ref{eq:u_rad}. 
We set the slope of the power-law internal energy distribution to $m\equiv -\mathrm{d}\ln u_\mathrm{rad}/\mathrm{d}\ln r=5$, which matches the radial pressure distributions in Figure \ref{fig:radial}, rather than the constant kinetic-to-internal energy ratio case (i.e., equipartition between kinetic and radiation energies at each layer) with $m=3$. 
In the cocoon breakout, a part of the CSM close to the jet axis is swept by the forward shock earlier and then accelerated to higher velocities, forming the outermost layer of the ejecta in the free-expansion stage. 
On the other hand, inner layers of the ejecta originated from CSM material located farther from the axis, which must have been hit by the shock later. While the internal energy of the shocked material would be comparable to the kinetic energy immediately after the shock passage, the ejected gas suffers from adiabatic cooling due to rapid expansion both in radial and lateral directions, leading to the freeze-out of the density structure. 
The faster layers in the ejecta should have experienced the density freeze-out earlier. 
This difference would make the internal energy distribution steeper than the kinetic energy distribution. 
We define the time $t_\mathrm{fr}$ at which the innermost layer experiences the freeze-out and assume that the spherical ejecta starts expanding at this freeze-out time $t=t_\mathrm{fr}$. 
The internal to kinetic energy ratio at the innermost layer is given by $f_\mathrm{th}=0.5$ at $t=t_\mathrm{fr}$. 
This freeze-out time is related with the breakout time of the jet from the CSM (see below). 

The expanding ejecta becomes dilute and transparent with time. 
Initially, photons in a radial layer are locked up in the layer and advected according to the local flow velocity due to the strong gas-radiation coupling via scattering. 
They start diffusing through layers when the diffusion velocity of radiation exceeds the local flow velocity (radiation breakout). 
The photons eventually reach the photosphere, above which they can travel freely. 
For a given constant and uniform opacity $\kappa$, which we assume to be $0.1\ \mathrm{cm}^2\,\mathrm{g}^{-1}$, we identify the breakout layer and photosphere. 
The trajectory of radiation initially trapped in each radial layer is numerically calculated and used to obtain the observed bolometric luminosity. 
The details are found in Appendix \ref{sec:light_curve}.

In the following, we assume a power-law ejecta with $M_\mathrm{ej}=0.1\,M_\odot$, and 
the kinetic energy of the whole ejecta is taken as $E_\mathrm{kin}=4.3\times 10^{51}\,\mathrm{erg}$. %
The freeze-out times are assumed to be $t_\mathrm{fr}=10,\ 10^2,\ 10^3,$ and $10^4\,\mathrm{s}$. 
The freeze-out time is defined so that the innermost layer with the velocity $c\beta_\mathrm{min}$ is in the free expansion regime. 
This is expected to happen when the ejected materials with $c\beta_\mathrm{min}$ travel at a distance comparable to $R_\mathrm{csm}$, i.e., when its radius doubles from the initial value of $R_\mathrm{csm}$. 
We note that a jet in a CSM with a larger outer radius emerges from the CSM surface at late epochs, and thus the density freeze-out happens later (i.e., the value of $t_\mathrm{fr}$ is larger for larger $R_\mathrm{CSM}$); 
the four different freeze-out times above roughly correspond to the CSM outer radii of $\simeq c\beta_\mathrm{min}t_\mathrm{fr}=3\times 10^{10}$, $3\times 10^{11}$, $3\times 10^{12}$, and $3\times 10^{13}\,\mathrm{cm}$, or $\simeq 0.4$, $4$, $40$, and $400\,R_\odot$, respectively, for $\beta_\mathrm{min}=0.1$.

In Figure \ref{fig:emission}, we show the temporal evolution of the radiative properties of the ejecta for different ejecta freeze-out times. 
The photospheric luminosity rises to the peak value within $t_\mathrm{obs}<0.1$ day and then decays with $L_\mathrm{ph}\propto t_\mathrm{obs}^{-1}$.
This behavior is a consequence of the recession of the breakout layer and the photosphere. 
The luminosity peak is indeed accompanied by the decrease in the velocity of the breakout layer. 
Models with earlier freeze-out times $t_\mathrm{fr}$ are characterized by lower luminosities. 
As we have already mentioned above, this is due to the adiabatic cooling of the ejecta. 
The epoch at which each layer in the ejecta becomes transparent is determined by the dynamical properties of the ejecta and is independent of the ejecta freeze-out time. 
The radiation energy in the ejecta, however, evolves as $(t/t_\mathrm{fr})^{-1}$. 
With an early freeze-out time and thus a small formation radius, the ejecta would lose a considerable amount of radiation energy due to expansion before becoming transparent and releasing thermal photons. 
On the other hand, the ejecta can stay hot with a late freeze-out epoch and thus a large formation radius.  

The ejecta releases all trapped photons until $t_\mathrm{obs}\simeq 1\,\mathrm{days}$, i.e., $\beta_\mathrm{br}=\beta_\mathrm{min}=0.1$, at which we stop calculations. 
In reality, emission from the inner non-relativistic ejecta powered by radioactive decay would also contribute to the luminosity at $t_\mathrm{obs}\gtrsim 1\,\mathrm{days}$, leading to the radioactive bump. 
Still, the light curves within a day should be considered reliable. 

\subsection{UV-optical flash}
\begin{figure}
\begin{center}
\includegraphics[scale=0.65]{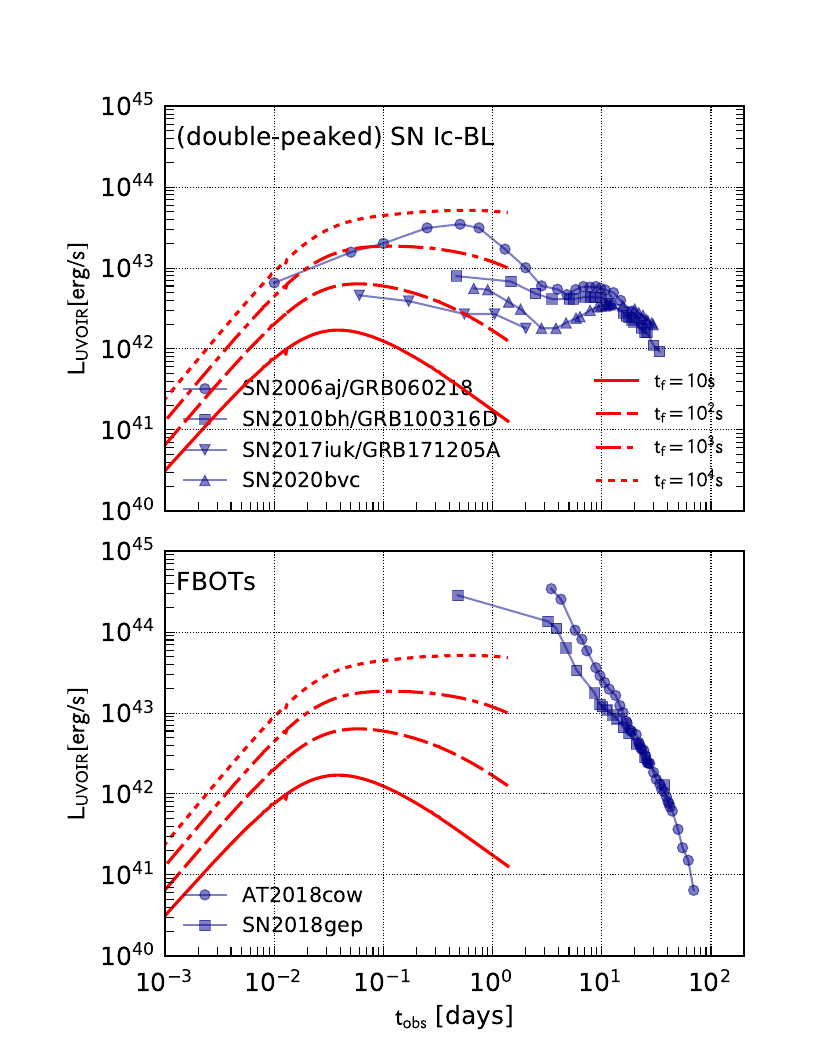}
\end{center}
\caption{Theoretical UVOIR light curves compared with some double-peaked broad-lined SNe-Ic (upper) and FBOTs (lower in the literature). 
The same model light curves are shown in both panels. 
In the upper panel, we plotted the light curves of SN~2006aj, 2010bh, 2017iuk, and 2020bvc. 
In the lower panel, we plotted the light curves of AT 2018cow and SN~2018gep. 
The light curve data are collected from \protect\citet{2011ApJ...740...41C} (SN~2006aj and 2010bh; their Figure 6), \protect\citet{2019Natur.565..324I} (SN~2017iuk), \protect\citet{2022ApJ...932..116H} (SN~2020bvc), \protect\citet{2019MNRAS.484.1031P} (AT 2018cow), and \protect\citet{2019ApJ...887..169H} (SN~2018gep).
}
\label{fig:lc_uvoir}
\end{figure}

In the bottom panel of Figure \ref{fig:emission}, we show the evolution of the effective temperature obtained by assuming the Stefan-Botlzmann law, Equation \ref{eq:Teff}. 
We note that the effective temperature estimated in this way is not necessarily consistent with the colour temperature, especially when the matter is scattering-dominated. 
We need more detailed radiative transfer simulations for a more accurate colour temperature estimation. 
As we discuss in Appendix \ref{sec:light_curve}, radiation equilibrium is not fully guaranteed at the outermost layer in models with long freeze-out times $t_\mathrm{fr}\gtrsim 10^3\,\mathrm{s}$. 
Therefore, in the beginning, the spectrum of the photoshperic emission could deviate from a Planck spectrum with the local gas temperature. 
This is especially true for photons staying in the original layer only within a period shorter than the assumed freeze-out time. 
The photospheric photons emitted at later epochs are from deeper layers and more easily achieve radiation equilibrium even though only the free-free process is at work for photon creation. 

Nevertheless, the effective temperature gives us a rough idea about the characteristic wavelength range of the photospheric emission. 
Our models show that the effective temperature can be initially as high as $T_\mathrm{eff}\sim10^5$--$ 10^6\,\mathrm{K}$ or $0.01$--$0.1\,\mathrm{keV}$, i.e., in the soft X-ray or EUV energy range, which is followed by a monotonic decrease to lower values. 
An effective temperature of $T_\mathrm{eff}\sim 10^4$--a few $10^4\,\mathrm{K}$ is expected around $t_\mathrm{obs}\sim 0.1$--$1\,\mathrm{days}$. 
Therefore, it is predicted that the cooling emission gives rise to bright UV/optical flash within $0.1$-$1\,\mathrm{days}$ after the explosion. 

We assume that the spectra of the photospheric emission are approximated as a Planck function with the temperature given by $T_\mathrm{eff}$ at each epoch. 
We then calculate UVOIR light curves by integrating the Planck function with respect to the wavelength from $2\times 10^2\,\mathrm{nm}$ to $2\times 10^4\,\mathrm{nm}$. 
The resultant UVOIR light curves are presented in Figure \ref{fig:lc_uvoir}. 
The UVOIR light curves reach the peak luminosity of $10^{42}$--$10^{44}\,\mathrm{erg}\,\mathrm{s}^{-1}$ at $t_\mathrm{obs}\simeq 0.03$--$0.1\,\mathrm{days}$, suggesting an hours-long bright UV-optical flash.  

The expected emission is brighter in UVOIR and more short-lived than the cooling envelope emission for normal SN explosions from RSG progenitors (e.g., \cite{2010ApJ...725..904N,2013ApJ...769...67P,2014ApJ...788..193N}). 
This is due to the relatively small ejecta mass, $\sim 0.1\,M_\odot$ and faster velocity $\sim 0.1c$. 
On the other hand, the expected photospheric temperature is more or less similar around the UVOIR peak, $T_\mathrm{ph}=10^4$--$10^5\,\mathrm{K}$. 
We discuss potential applications of this ejecta cooling emission for transient populations in Section \ref{sec:discussion}.

\section{Discussion}\label{sec:discussion}
In this section, we discuss observational imprints of the ejecta produced by jet-CSM interaction. 
We focus on GRB-SNe and FBOTs as potential astronomical transients powered by jet-CSM interaction. 

\subsection{Energetic SNe and GRBs}
The cooling emission from the mildly relativistic ejecta is naturally expected in CCSNe associated with GRBs. 
In on-axis events, however, the bright synchrotron afterglow emission from the relativistic jet would overwhelm the thermal emission. 
The thermal component may instead be clearly detected in events with intrinsically weak and/or off-axis jets or choked jets. 
Even without association with apparent GRBs, some energetic CCSNe could harbor GRB-like engines, which also leads to the creation of mildly relativistic ejecta. 

\subsubsection{GRB-SNe/SNe Ic-BL with early UV bump}
Some nearby GRB-SNe, such as GRB~060218/SN~2006aj \citep{2006Natur.442.1008C,2006Natur.442.1018M,2006ApJ...645L..21M,2006Natur.442.1014S}, GRB~100316D/SN~2010bh\citep{2011MNRAS.411.2792S,2011ApJ...740...41C,2012ApJ...753...67B,2012A&A...539A..76O}, and GRB~171205A/SN~2017iuk \citep{2018A&A...619A..66D,2018ApJ...867..147W,2019Natur.565..324I}, certainly exhibited bright UV emission prior to the main SN light powered by $^{56}$Ni (``double-peaked'' light curves). 
These nearby events are classified into a low-luminosity class of GRBs. 
It is still unclear whether these events are driven by a weak jet or more spherical shock breakout (e.g., \cite{2016MNRAS.460.1680I}). 

In Figure \ref{fig:lc_uvoir}, we compare the UVOIR light curves of SNe~2006aj, 2010bh, and 2017iuk with our theoretical light curve models. 
For SN~2017iuk, we plot the luminosity of the blackbody component provided by \citet{2019Natur.565..324I} (their Extended Data Table 2). 
SN~2006aj shows the brightest early UV bump among these three events. 
The peak luminosity of $\sim3\times 10^{43}\,\mathrm{erg}\,\mathrm{s}^{-1}$ is similar to the model with $t_\mathrm{fr}=10^3\,\mathrm{s}$, but the observed peak is at $\sim 0.5\,\mathrm{day}$ after the GRB trigger rather than the theoretical peak epoch of $<0.1\,\mathrm{day}$. 
The other two GRB-SNe show less luminous and monotonically decaying UVOIR luminosity at similar epochs of $0.1$--$1\,\mathrm{day}$. 
The required short ejecta freeze-out time $t_\mathrm{fr}$ suggests dense materials with outer radii smaller than $400\,R_\odot$. 
Since the required radii are within typical BSG and RSG radii, such dense materials may be realized as a tenuous envelope still attached to the progenitor star. 
An intense mass-loss immediately prior to the explosion is still another promising way to produce such confined CSMs. 
For a typical wind velocity of Wolf-Rayet stars, $v_\mathrm{wind}\sim 10^3\,\mathrm{km}\,\mathrm{s}^{-1}$, such mass-loss must have happened only $R_\mathrm{out}/v_\mathrm{wind}\sim 0.3$--$3\,\mathrm{days}$ before the gravitational collapse, which corresponds to silicon burning stage for massive progenitors.

Recent optical surveys have successfully detected some SNe Ic-BL as early as $<1$ day after the estimated explosion dates. 
SN~2020bvc \citep{2020ApJ...895...49H,2020A&A...639L..11I,2021ApJ...908..232R} is among the best studied object and exhibited a clear signature of early UV emission. 
The early UV bump resembles those of GRB-SNe as shown in the upper panel of Figure \ref{fig:lc_uvoir}, and thus can be explained by the cooling emission. 
An alternative possibility is the interaction-powered emission from spherical SN ejecta colliding with a spherical CSM as demonstrated by the multi-band light curve model of \citet{2021ApJ...910...68J}. 
However, the suggested collision with a spherical $\sim 0.1\,M_\odot$ CSM would have suppressed the outer high-velocity part of the ejecta, which is inconsistent with the presence of high-velocity absorption features identified in the early spectra of this object and likely caused by heavy elements \citep{2020ApJ...895...49H,2020A&A...639L..11I,2023MNRAS.522.2267M}. 
Therefore, we conclude that the early UV emission from SN Ic-BL 2020bvc is plausibly powered by a jet or outflow penetrating an extended material. 

\subsubsection{Relation to off-axis GRBs}
As we have argued above, low-luminosity GRBs with early UV thermal emission can be explained by GRB jets interacting with massive CSM. 
On the other hand, successful GRB jets seen from off-axis viewing angles should also be observed as under-luminous events and can potentially constitute a fraction of low-luminosity GRBs (e.g., \cite{2002ApJ...571L..31Y,2003ApJ...594L..79Y}). 
For a successful GRB jet from a compact star, the associated formation of quasi-spherical ejecta should happen at radii of the order of $\sim R_\odot$ (the corresponding freeze-out time of a few 10 seconds). 
This leads to only a dim early UV thermal emission due to adiabatic cooling \citep{2017ApJ...834...28N}. 
The UVOIR light curve with $t_\mathrm{fr}=10\,\mathrm{s}$ in Figure \ref{fig:lc_uvoir} indicates the peak luminosity of $\sim 2\times 10^{42}\,\mathrm{erg}\,\mathrm{s}^{-1}$ at $t_\mathrm{obs}\sim 0.04\,\mathrm{days}$ for such small formation radii. 
The emission should also be overwhelmed by the optical afterglow synchrotron emission from the off-axis jet. 
More quantitative discussion on how the off-axis afterglow emission hides the early thermal emission requires off-axis afterglow modelling, which we leave as a future work. 
Nevertheless, the absence of bright early thermal UV emission in a low-luminosity GRB implies the absence of a massive and extended CSM causing the jet dissipation and can distinguish genuine off-axis jets from CSM-powered events.

\subsubsection{Ultra-long GRBs}
Some GRBs with unusually long burst duration ($\gtrsim$several $10^3\,\mathrm{s}$) are classified as ultra-long GRBs (e.g., \cite{2014ApJ...781...13L}). 
Although the associated SN component was faint, the ultra-long GRB~101225A \citep{2011Natur.480...69C,2011Natur.480...72T} shares similar X-ray and UV emission properties with nearby low-luminosity GRBs. 
GRB~111209A is another ultra-long GRB found to be associated with a luminous SN component (SN~2011kl; \cite{2015Natur.523..189G,2019A&A...624A.143K}). 
Although their origin is still unclear, their long burst duration may indicate an engine powered by a long-lasting accretion of stellar materials on to a central compact object and/or the dissipation of the jet energy at unusually large radii. 
Massive stars with more extended envelopes than typical Wolf-Rayet stars, in particular, low-metallicity BSGs or pop III stars are suggested to be a possible progenitor scenario for ultra-long GRBs (e.g., \cite{2012ApJ...759..128N,2013ApJ...778...67N,2013ApJ...766...30G}). 

Our simulation with $10\,M_\odot$ CSM (an extended envelope in this case) with $R_\mathrm{out}=40\,R_\odot$ would imitate such progenitors. 
The corresponding UVOIR light curve models (the ejecta freeze-out time of $10^3\,\mathrm{s}$ or shorter) predict a peak luminosity up to $\sim 10^{43}\,\mathrm{erg}\,\mathrm{s}^{-1}$ in $\sim 0.1$ day. 
This UV emission would be deeply buried in a bright optical afterglow and difficult to identify. 
However, confirming the presence/absence of this emission component arising from the accompanied cocoon can be an important observational test for the BSG scenario for ultra-long GRBs.

\subsubsection{Impact on the early optical spectra}
Even when the bright cooling emission is somehow suppressed, the mildly relativistic ejecta can manifest itself in the early optical spectra of any potential transients harboring jet activity. 
The recent spectral synthesis study by \citet{2023MNRAS.522.2267M} demonstrates that energetic CCSNe with and without sub-relativistic outer layers can be distinguished by early optical spectra observed within $\sim 7\,$days. 
The early spectra lack sharp absorption/emission features due to line broadening, but are sensitive to the extent of chemical mixing and the outermost ejecta velocity. 
Although they adopted slightly steeper density slope of $\mathrm{d}\ln\rho/\mathrm{d}\ln v=-6$ than $\mathrm{d}\ln\rho/\mathrm{d}\ln v=-5$ suggested by our simulations, the synthetic spectra show a remarkable agreement with the early spectra of SNe~2017iuk and 2020bvc. 
This finding suggests the power-law ejecta created by the jet-CSM interaction can be a promising mechanism to explain spectral features of some, if not all, energetic SNe in their infant phase. 
The universal density structure of the mildly relativistic ejecta revealed by \citet{2022ApJ...925..148S} and this work strongly suggests that the presence of such ejecta would be quite common among GRB-SNe and SNe Ic-BL powered by long-lasting jet activity.

\subsection{Potential link to FBOTs}
Another population of (potentially) jet-powered optical transients is FBOTs. 
Some radio-luminous FBOTs are likely powered by $0.1\,M_\odot$ ejecta traveling at $\sim0.1c$ \citep{2020ApJ...895L..23C,2022ApJ...926..112B,2022ApJ...932..116H}. 
In addition, some SN Ic-BL with heavily UV-dominated early emission, such as iPTF16asu \citep{2017ApJ...851..107W} and SN~2018gep \citep{2019ApJ...871...73H,2021ApJ...915..121P}, may constitute a part of the FBOTs class. 

The suggested ejecta properties of FBOTs are consistent with those of the ejecta created from jet-CSM interaction as demonstrated by our simulations. 
In the lower panel of Figure \ref{fig:lc_uvoir}, we compare the theoretical UVOIR light curves with AT 2018cow and SN~2018gep. 
We note that we only plot the luminosity at epochs with UV photometry available for SN~2018gep (see, \cite{2019ApJ...871...73H}). 
The initial luminosity of SN~2018gep is more luminous than our UVOIR light curve model with the longest $t_\mathrm{fr}=10^4\,\mathrm{s}$. 
Creating hot ejecta at sufficiently late epochs, $t_\mathrm{fr}>10^4\,\mathrm{s}$, requires a highly extended CSM with an outer radius larger than $R_\mathrm{csm}>400\,R_\odot$. 
The more luminous AT 2018cow requires an even more extended CSM. 
Such extended CSMs are unlikely to be stellar envelopes bound to the progenitor star and may be distinguished from compact CSMs/envelopes required for the double-peaked SNe Ic-BL. 

Recently, \citet{2022MNRAS.513.3810G} proposed a similar cooling ejecta (or cocoon) scenario for luminous FBOTs like AT 2018cow. 
They treat the thermal emission from the cooling ejecta by blackbody radiation from the photosphere, i.e., they identify the photosphere in a hydrodynamic profile and assume a blackbody flux density $\sigma T^4$ with the local temperature $T$. 
This treatment requires some fast photon creation process that immediately supplies fresh photons in the photospheric layer and always maintains thermal equilibrium there, and thus can overestimate the photospheric luminosity. 
On the other hand, we consider the energy release from radial layers according to the recession of the photosphere. 
The photons lost through the photosphere are not replenished except for those diffusing through from deeper layers. 
This leads to less luminous cooling emission than the model presented by \citet{2022MNRAS.513.3810G}.  

We remark on a couple of difficulties of cocoon/ejecta models for AT~2018cow.  
First, its almost monotonically declining photospheric radius evolution seems inconsistent with normal freely expanding ejecta \citep{2020ApJ...897..156U}. 
Second, AT~2018cow showed significant X-ray emission characterized by a complicated X-ray spectrum, which cannot be explained by our simple cooling emission model \citep{2019ApJ...872...18M}.

\section{Summary}\label{sec:summary}
In this work, we conduct a series of GRB jet simulations with a $14M_\odot$ CO star surrounded by a massive CSM. 
After penetrating the progenitor, the jet propagates in the dense ambient medium and a fraction of the shocked gas is ejected into outer space with relativistic velocities exceeding $0.1c$. 
We aim at clarifying the dynamical properties of the mildly relativistic ejecta produced by the jet-CSM interaction which provides a basis for modelling the electromagnetic signals from any astronomical transients powered by relativistic jets. 
Our findings are summarized as follows; 
(1) The jet morphology makes a transition from a cold jet to a choked jet outflow driven by thermal pressure. 
For the CSM outer radii explored in this work ($40R_\odot$ and $400R_\odot$), most models are in the intermediate stage, in which the jet material is swept by the reverse shock formed as a result of the mass-loading, but still the forward shock is faster along the jet axis.  
(2) A denser CSM more effectively confines the jet, thereby making the volume of the shocked region small. 
As a result, the mass of the materials accelerated beyond $0.1c$ does not linearly scale with the CSM mass. 
(3) The radial density structure of the mildly relativistic ejecta is remarkably universal and well represented by a power-law function of the radial velocity with an exponent of $\simeq -5$, $\rho\propto r^{-5}$, which is flatter than normal CCSN ejecta resulting from an almost point explosion at the center of the progenitor. 
This finding reconfirms our earlier results \citep{2022ApJ...925..148S} and extends the applicability of the ejecta with the flat density structure for wider parameter ranges. 
(4) The mass and the kinetic energy of material accelerated beyond $0.1c$ are not linearly scaled with the CSM mass, resulting in relatively uniform properties for a wide range of the CSM mass and radius; the jet properties adopted in this work ($5 \times 10^{51}$ erg with the duration of 20s) represent typical GRB-jet properties, and in this case we find that the resulting relativistic component has $\simeq0.05$--$0.12M_\odot$ and $\simeq1$--$5\times10^{51}\,\mathrm{erg}$. 

The fast $\sim 0.1M_\odot$ ejecta resulting from the jet-CSM interaction is predicted to give rise to a bright UV-optical flash powered by the remaining thermal energy in the ejecta (i.e., the shock cooling emission). 
The UVOIR light curves are characterized by a peak luminosity of $10^{42}$--$10^{44}\,\mathrm{erg}\,\mathrm{s}^{-1}$ and a duration of $0.1$--$1$ days. 
The observational counterparts of this ejecta cooling emission could be GRB-SNe/SNe Ic-BL in their infant stages or luminous FBOTs. 
Also, a fraction of the kinetic energy of the ejecta would be eventually converted into non-thermal radiation by the collision with the ambient gas followed by high-energy particle acceleration. 
Future more detailed computations of the expected light curve and spectral synthesis are highly encouraged to unveil essential roles played by jet-powered events in the transient universe.

\begin{ack}
The authors appreciate the anonymous referee for his/her careful reading of the manuscript and comments. 
A.S. acknowledges support by Japan Society for the Promotion of Science (JSPS) KAKENHI Grant Number JP19K14770 and JP22K03690.  CI is supported by the JST FOREST Program (JPMJFR2136) and the JSPS Grant-in-Aid for Scientific Research (23H01169).  K. M. is supported by JSPS KAKENHI Grant Number JP20H00174 and JP24H01810. 
The computations in this work are partly performed on Resceubbc at the Research Center for the Early Universe, The University of Tokyo, and XC50 system at National Astronomical Observatory of Japan. 
\end{ack}

\appendix 

\section{Thermal emission model}\label{sec:light_curve}
In this section, we describe the light curve model for the thermal emission from a cooling ejecta. 
The thermal photons trapped in each radial layer diffuse through the ejecta and are eventually released into outer space. 
An important assumption is that thermal photons simply diffuse outwardly and are not replenished once they escape the layer. 
This condition is met for scattering-dominated media at high temperature. 
We neglect some relativistic effects $(\Gamma\simeq 1)$ for keeping some of the the following computations analytically tractable. 
We plan to carry out radiative transfer simulations incorporating relativistic effects and non-uniform opacity in future work. 

\subsection{Power-law ejecta}
First of all, we specify the ejecta structure. 
We consider a spherical freely expanding gas with the radial density profile given by
\begin{equation}
\rho(t,r)=\rho_0\left(\frac{t}{t_\mathrm{fr}}\right)^{-3}
\left(\frac{\beta}{\beta_\mathrm{min}}\right)^{-n}.
\label{eq:density}
\end{equation}
for the (normalized) velocity $\beta=r/(ct)$ between the minimum and maximum velocities $\beta_\mathrm{min}$ and $\beta_\mathrm{max}$. 
We assume that the ejecta forms and starts expanding homologously at $t=t_\mathrm{fr}$, which corresponds to the freeze-out of the density structure for the innermost layer. 
The index $n$ should be larger than $3$ to keep the ejecta mass finite. 
In the following, we fix $n=5$ in accordance with the main text (Sec. \ref{sec:averaged_radial_profiles}). 
For simplicity, we assume no medium both inside and outside the ejecta. 
The normalization constant is given as a function of the mass $M_\mathrm{ej}$ as
\begin{equation}
\rho_0=\frac{(n-3)M_\mathrm{ej}}
{4\pi(c\beta_\mathrm{min}t_\mathrm{fr})^3}
\left[1-\left(\frac{\beta_\mathrm{max}}{\beta_\mathrm{min}}\right)^{3-n}\right]^{-1}. 
\end{equation}

We assume that the internal energy of the ejecta is dominated by radiation. 
The radiation energy density of the ejecta is also assumed to follow a power-law profile. 
We specify the radial distribution with a power-law index $m$ and a normalization parameter $f_\mathrm{th}$;
\begin{equation}
    u_\mathrm{rad}(t,\beta)=f_\mathrm{th}\rho_0c^2\frac{\Gamma_\mathrm{min}^3\beta_\mathrm{min}^2}{\Gamma_\mathrm{min}+1}
    \left(\frac{t}{t_\mathrm{fr}}\right)^{-4}
    \left(\frac{\beta}{\beta_\mathrm{min}}\right)^{-m},
\end{equation}
where $\Gamma_\mathrm{min}$ is the Lorentz factor corresponding to the minimum velocity $\beta_\mathrm{min}$. 
At the innermost layer with $\beta=\beta_\mathrm{min}$, the ratio of the radiation energy to the kinetic energy is initially given by $f_\mathrm{th}$ ($t=t_\mathrm{fr}$). 
We adopt a fixed value of $f_\mathrm{th}=0.5$ in the following. 
The power-law index $m=3$ gives the uniform radiation-to-kinetic energy ratio $f_\mathrm{th}$ throughout the entire ejecta. 
The radial pressure distributions in Figure \ref{fig:radial}, however, show steeper slopes. 
This implies that faster layers experience the density freeze-out earlier, resulting in a longer period for adiabatic cooling than slower layers. 
We examine the case with $m=5$ in the following. 

\subsection{Radiation equilibrium}
Here we examine whether the ejecta is in radiation equilibrium. 
With $(\beta_\mathrm{min},\beta_\mathrm{max})=(0.1,0.7)$ used in the main text, the initial density at the outermost ejecta is calculated as
\begin{equation}    \rho(t_\mathrm{fr},c\beta_\mathrm{max}t_\mathrm{fr})=7.0\times 10^{-11}
\left(\frac{M_\mathrm{ej}}{0.1\,M_\odot}\right)
\left(\frac{t_\mathrm{fr}}{10^3\,\mathrm{s}}\right)^{-3}
\,\mathrm{g}\,\mathrm{cm}^{-3}
\end{equation}
with the corresponding initial internal energy given by
\begin{eqnarray}
    u_\mathrm{rad}(t_\mathrm{fr},c\beta_\mathrm{max}t_\mathrm{fr})
    &=&1.6\times 10^{8}
    \left(\frac{f_\mathrm{th}}{0.5}\right)\nonumber\\
    &&\times
    \left(\frac{M_\mathrm{ej}}{0.1\,M_\odot}\right)
    \left(\frac{t_\mathrm{fr}}{10^3\,\mathrm{s}}\right)^{-3}
    \,\mathrm{erg}\,\mathrm{cm}^{-3}.
    \label{eq:u_rad}
\end{eqnarray}
Assuming the expected equilibrium temperature of 
\begin{equation}
    T_\mathrm{eq}=\left[\frac{u_\mathrm{rad}}{a_\mathrm{r}}\right]^{1/4}\simeq 3.8\times 10^5
    \left(\frac{u_\mathrm{rad}}{1.6\times 10^8\,\mathrm{erg}\,\mathrm{cm}^{-3}}\right)^{1/4}
    \,\mathrm{K},
\end{equation}
the free-free emission with the emissivity $\epsilon_\mathrm{ff}$ can maintain the radiation energy density of
\begin{eqnarray}
    \epsilon_\mathrm{ff}t_\mathrm{fr}&\simeq& 2.5\times 10^7
    \left(\frac{T}{3.8\times10^5\,\mathrm{K}}\right)^{1/2}
    \\&&\times
    \left(\frac{\rho}{7.0\times10^{-11}\,\mathrm{g}\,\mathrm{cm}^{-3}}\right)^2
    \left(\frac{t_\mathrm{fr}}{10^3\,\mathrm{s}}\right)
    \,\mathrm{erg}\,\mathrm{cm}^{-3},
    \nonumber
\end{eqnarray}
for $t_\mathrm{fr}=10^3\,\mathrm{s}$ at the outermost layer. 
Here we assume fully ionized gas with pure oxygen composition and a Gaunt factor of unity. 
With $t_\mathrm{fr}=10^3\,\mathrm{s}$, for example, the ratio $\epsilon_\mathrm{ff}t_\mathrm{fr}/u_\mathrm{rad}<1$ smaller than unity means that free-free emission alone cannot produce enough photons to maintain the radiation equilibrium. 
Therefore, at least around the outermost layers, the energy distribution of photons can deviate from the Planck spectrum with the local gas temperature. 
Because of the strong density dependence of the free-free emissivity, models with shorter freeze-out times, $t_\mathrm{fr}=10$ and $10^2\,\mathrm{s}$, safely achieve the radiation equilibrium. 
With $t_\mathrm{fr}=10^4\,\mathrm{s}$, on the other hand, the outermost layer more seriously suffers from photon paucity. 

However, due to the steep density increase toward the inner boundary, inner layers more easily achieve the radiation equilibrium within shorter time-scales. 
With $t_\mathrm{fr}=10^3\,\mathrm{s}$, for example, the layer with $\beta<0.5$ achieves the radiation equilibrium within $t=t_\mathrm{fr}$. 
In Section \ref{sec:radiation}, we focus on emission around $t_\mathrm{obs}=10^{-2}$--$1\,\mathrm{days}$, at which photospheric emission is dominated by UV/optical photons. 
It is coincident with the photosphere receding into deeper layers. 
This makes the assumption of radiation equilibrium and the colour temperature determined by the Stefan-Boltzmann law more reliable in the decaying phase of the luminosity evolution (e.g., $t_\mathrm{obs}>10^{-2}\,\mathrm{days}$). 

Another remark is that some layers close to the outer edge are already optically thin at the beginning $t=t_\mathrm{fr}$ and therefore the photons could be far from equilibrium. 
However, the mass of the initially optically-thin layer is only a tiny fraction of the total mass. 
As shown in the velocity evolution in Figure \ref{fig:emission} (the 2nd panel), the photosphere and the breakout layer (defined in Section A3 below) are almost identical at the beginning and recede into the ejecta only after $t_\mathrm{obs}=0.1\,\mathrm{days}$. 
The photons located in the initially optically-thin layers therefore constitute only a small fraction of the total radiation emitted by the ejecta. 
\begin{figure}
\begin{center}
\includegraphics[scale=0.55]{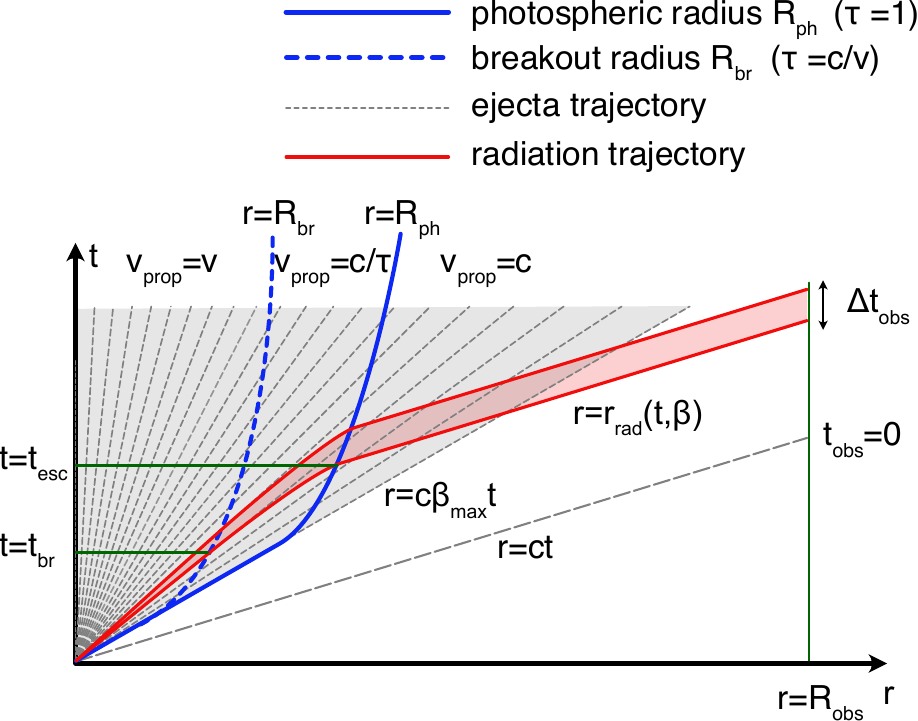}
\end{center}
\caption{Space-time diagram for the ejecta expansion and the radiation propagation. 
}
\label{fig:schematic}
\end{figure}
\subsection{Radiation breakout and escape}
Next, we introduce a couple of characteristic radii in the ejecta; the breakout layer and the photosphere. 
The breakout layer is by definition more deeply located than the photosphere (see below). 
These two boundaries thus divide the ejecta into three different regions in terms of gas-radiation coupling. 
Figure \ref{fig:schematic} schematically represents the ejecta expansion and the propagation of radiation in a space-time diagram. 
When photons are trapped in a dense radial layer, the strong gas-radiation coupling via electron scattering does not allow them to diffuse across different layers. 
In this strong-coupling limit, the photons are simply advected along the gas flow and propagate at the local velocity. 
When the layer becomes more dilute, the photons are now able to diffuse through radial layers outwardly while they are still coupled with gas via scattering. 
The photons eventually reach the photosphere, beyond which they travel almost freely at the speed of light. 

In practice, the breakout layer and the photosphere are defined in the following way. 
We consider a radially traveling photon ray from radius $r$ at $t$ (the corresponding velocity coordinate is therefore $\beta=r/(ct)$). 
The ejecta keeps expanding while the ray is traveling through the ejecta. 
Since the position of the photon at $t'(>t)$ is given by $r+c(t'-t)$, the optical depth along the radial ray to the surface of the ejecta is calculated as follows,
\begin{eqnarray}
\hspace{-1em}
\tau(t,r)&=&\int^{t_\mathrm{out}}_t\kappa\rho(t',r+c(t'-t))\mathrm{d} (ct')
\nonumber\\&=&
\kappa\rho_0ct_\mathrm{fr}\beta_\mathrm{min}^n\left(\frac{t}{t_\mathrm{fr}}\right)^{-2}
\\&&\times
\frac{\beta^{1-n}\left[n-2-(n-1)\beta\right]-
\beta_\mathrm{max}^{1-n}\left[n-2-(n-1)\beta_\mathrm{max}\right]}{(n-2)(n-1)(1-\beta)^2},
\nonumber
\end{eqnarray}
for a constant opacity $\kappa$. 
Note that the photon ray reaches the outermost layer at $t_\mathrm{out}=(t-r/c)/(1-\beta_\mathrm{max})$. 

The photosphere at $t$ is defined as the radial layer ($r=R_\mathrm{ph}=c\beta_\mathrm{ph}t$) satisfying $\tau(t,R_\mathrm{ph})=1$. 
At the breakout layer ($r=R_\mathrm{br}=c\beta_\mathrm{br}t$), the photon diffusion velocity $c/\tau(t,R_\mathrm{br})$ is equal to the radial velocity of the layer, $c/\tau(t,R_\mathrm{br})=v$, which is rewritten as $\tau(t,R_\mathrm{br})=1/\beta_\mathrm{br}$. 
The breakout and photospheric radius as a function of time $t$ are therefore numerically obtained by solving these non-linear equations $\tau(t,R_\mathrm{ph})=1$ and $\tau(t,R_\mathrm{br})=1/\beta_\mathrm{br}$ with some root-finding technique. 

\subsection{Radiation trajectory}
We consider the trajectory of radiation initially trapped at a layer with the velocity coordinate $\beta$. 
As is schematically represented in Figure \ref{fig:schematic}, the propagation of radiation is divided into the three regimes. 
The radial propagation velocity of radiation $v_\mathrm{prop}$ is given as a function of time $t$ and the radial coordinate $r$,
\begin{equation}
    v_\mathrm{prop}(t,r)=\left\{
    \begin{array}{ccl}
    r/t&\mathrm{for}&r\leq R_\mathrm{br}(t),\\
    \frac{c}{\tau(t,r)}&\mathrm{for}&
    R_\mathrm{br}(t)<r\leq R_\mathrm{ph}(t),\\
    c&\mathrm{for}&R_\mathrm{ph}(t)<r.
    \end{array}
    \right.
\end{equation}
The radiation trajectory as a function of $t$ is computed by integrating this propagation velocity,
\begin{equation}
    r_\mathrm{rad}(t,\beta)=\int^t_0 v_\mathrm{prop}(t',r_\mathrm{rad})dt'.
\end{equation}
In the advection regime, $r\leq R_\mathrm{br}(t)$, the radiation trajectory is a straight line, $r_\mathrm{rad}=c\beta t$, in the space-time diagram (Figure \ref{fig:schematic}). 
The radiation trajectory reaches the breakout layer at $t=t_\mathrm{br}$ and then enters the diffusion regime. 
In the diffusion regime, $R_\mathrm{br}(t)<r \leq R_\mathrm{ph}(t)$, the propagation velocity is dependent on the time and radial coordinates. 
The radiation trajectory in this regime is thus obtained by numerical integration. 
The diffusion velocity increases as the radiation propagates in the ejecta and approaches the speed of light. 
Therefore, the radiation trajectory gradually changes its slope as presented in Figure \ref{fig:schematic}. 
The radiation passes through the photosphere at $t=t_\mathrm{esc}$, at which the diffusion velocity is identical with the speed of light,  and then propagates freely in the remaining space. 
In the space-time diagram, the trajectory is again represented by a straight line but with the slope parallel to $r=ct$. 

\subsection{Luminosity}
Next, we calculate the radiation energy released by the recession of the characteristic radii in the velocity coordinate. 
We consider two adjacent radiation trajectories whose initial velocity coordinates are $\beta$ and $\beta-\Delta\beta$. 
As in Figure \ref{fig:schematic}, the radiation energy within the two layers is released into the surrounding free-propagation space (red shaded region). 
The radiation energy in this concentric shell at the emergence from the photosphere is approximately given by 
\begin{equation}
    \Delta E_\mathrm{rad}=
    u_\mathrm{rad}(t_\mathrm{fr},\beta)
    \Delta V_\mathrm{fr}
    \left(\frac{\Delta V_\mathrm{esc}}{\Delta V_\mathrm{fr}}\right)^{-1/3}
    ,
    \label{eq:dE}
\end{equation}
with
\begin{equation}
    \Delta V_\mathrm{fr}=
    4\pi c^3\beta^2\Delta \beta t_\mathrm{fr}^3,
\end{equation}
and
\begin{equation}
    \Delta V_\mathrm{esc}=\frac{4\pi}{3} 
    \left[r_\mathrm{rad}(t_\mathrm{esc},\beta)^3-
    r_\mathrm{rad}(t_\mathrm{esc},\beta-\Delta \beta)^3\right],
\end{equation}
being the volume of the shell at $t=t_\mathrm{fr}$ and $t_\mathrm{esc}$. 
The final term in Eq. (\ref{eq:dE}) therefore represents the adiabatic loss caused by the expansion of the shell from $t=t_\mathrm{fr}$ to $t_\mathrm{esc}$, while the remaining terms correspond to the initial radiation energy. 

The two radiation trajectories reach the observer at $r=R_\mathrm{obs}$ with a time delay $\Delta t_\mathrm{obs}$, which can be computed with the radiation trajectories. 
Since the radiation energy $\Delta E_\mathrm{rad}$ is received by the observer during this time interval, the observed luminosity is given by 
\begin{equation}
    L_\mathrm{obs}=\frac{\Delta E_\mathrm{rad}}{\Delta t_\mathrm{obs}}.
\end{equation}

\subsection{Temperature}
Finally, assuming the Stefan-Botlzmann law, the effective temperature $T_\mathrm{eff}$ is calculated as follows,
\begin{equation}
T_\mathrm{eff}=
\left(\frac{L_\mathrm{bol}}{4\pi\sigma r_\mathrm{ph}^2}\right)^{1/4}.
\label{eq:Teff}
\end{equation}



\end{document}